\documentclass[
  aps,
  prfluids,
  amsmath,
  amsfonts,
  amssymb,
  10pt,
  longbibliography,
  nofootinbib,
]{revtex4-2}

\usepackage[utf8]{inputenc}
\usepackage{microtype}
\usepackage{xparse}

\usepackage{booktabs}
\usepackage[hidelinks]{hyperref}
\usepackage{siunitx}
\usepackage{bm}
\usepackage[ISO]{diffcoeff}
\usepackage[normalem]{ulem}

\usepackage{graphicx}
\usepackage[svgnames]{xcolor}
\usepackage{csquotes}

\usepackage{newtxtext}
\usepackage[varvw]{newtxmath}

\newcommand*{\dd}{\text{d}}

\newcommand*{\cross}{\times}
\newcommand*{\gradient}{\bm{\nabla}}
\newcommand*{\divergence}{\gradient \cdot}
\newcommand*{\laplacian}{\nabla^2}

\NewDocumentCommand{\evec}{ o }{\bm{e}\IfValueTF{#1}{_{#1}}{}}

\newcommand*{\rms}{\text{rms}}
\newcommand*{\Emean}[1]{\langle #1 \rangle}  
\newcommand*{\Lmean}[1]{\langle #1 \rangle}  

\newcommand*{\xvec}{\bm{x}}
\newcommand*{\kvec}{\bm{k}}
\newcommand*{\rvec}{\bm{r}}
\newcommand*{\rvecUnit}{\hat{\rvec}}
\newcommand*{\kforcing}{k_\text{f}}
\newcommand*{\uvec}{\bm{u}}
\newcommand*{\Fvec}{\bm{F}}
\newcommand*{\vort}{\omega}
\newcommand*{\vortvec}{\bm{\vort}}

\newcommand*{\Rey}{\text{Re}}     
\newcommand*{\diss}{\varepsilon}  
\newcommand*{\taueta}{\tau_\eta}
\newcommand*{\lambdaTaylor}{\lambda}
\newcommand*{\ReLambda}{\Rey_{\lambdaTaylor}}
\newcommand*{\urms}{u_\rms}
\newcommand*{\vortrms}{\omega_\rms}
\newcommand*{\TL}{T_L}  

\newcommand*{\OmegaRot}{\Omega_{\text{rot}}}
\newcommand*{\RotationVec}{\bm{\Omega}_{\text{rot}}}
\newcommand*{\Rossby}{\text{Ro}}
\newcommand*{\RossbyL}{\Rossby_L}
\newcommand*{\RossbyOmega}{\Rossby_\omega}

\newcommand*{\Trot}{T_{\text{rot}}}

\newcommand*{\RoLarge}{0.165}
\newcommand*{\RoMid}{0.076}
\newcommand*{\RoSmall}{0.032}

\NewDocumentCommand{\xp}{ o }{\bm{X}\IfValueTF{#1}{_{#1}}{}}
\NewDocumentCommand{\vp}{ o }{\bm{v}\IfValueTF{#1}{^{(#1)}}{}}
\RenewDocumentCommand{\ap}{ o }{\bm{a}\IfValueTF{#1}{^{(#1)}}{}}  
\newcommand*{\rhovec}[1]{\bm{\rho}^{(#1)}}

\newcommand*{\gval}[1]{g_{#1}}
\newcommand*{\gvec}[1]{\bm{g}_{#1}}
\newcommand*{\InertiaTensor}{g}
\newcommand*{\corrvel}{\rho^{(v)}}
\newcommand*{\LambdaRegular}{\Lambda_\text{reg}}
\newcommand*{\AreaRegular}{A_\text{reg}}
\newcommand*{\Alignment}{\psi}

\newcommand*{\mat}[1]{\bm{\mathsf{#1}}}
\newcommand*{\InertiaTensorMat}{\mat{\InertiaTensor}}

\newcommand*{\gvalTriad}[1]{\gval{#1}^{\text{(triad)}}}
\newcommand*{\gvecTriad}[1]{\gvec{#1}^{\text{(triad)}}}
\newcommand*{\ITriad}[1]{I_{#1}^{\text{(triad)}}}

\newcommand*{\DeltaFull}{\Delta}  
\newcommand*{\DeltaIJ}[2]{\Delta_{#1}^{\! ({#2})}}
\newcommand*{\StwoIJ}[2]{S_{2, #1}^{({#2})}}

\newcommand*{\Rcomp}{\widetilde{R}}
\newcommand*{\Dist}{d}
\newcommand*{\DistSq}{\Dist^2}

\definecolor{forestgreen}{rgb}{0.13, 0.55, 0.13}

\begin{document}

\title{Multi-particle Lagrangian statistics in homogeneous rotating turbulence}
\author{Juan Ignacio Polanco}
\email{juan-ignacio.polanco@cnrs.fr}
\author{S.~Arun}
\author{Aurore Naso}
\email{aurore.naso@cnrs.fr}
\affiliation{Univ Lyon, CNRS, Ecole Centrale de Lyon, INSA Lyon, Univ Claude
  Bernard Lyon 1, Laboratoire de M\'ecanique des Fluides et d'Acoustique, UMR
  5509, 69130 Ecully, France
}

\begin{abstract}
Geophysical flows are often turbulent and subject to rotation.
This rotation modifies the structure of turbulence and is thereby expected to sensibly affect its Lagrangian properties.
Here, we investigate the relative dispersion and geometry of pairs, triads and tetrads in homogeneous rotating turbulence, by using direct numerical simulations at different rotation rates.
Pair dispersion is shown to be faster in the vertical direction (along the rotation axis) than in the horizontal one.
At long times, in Taylor's regime, this is due to the slower decorrelation of the vertical velocity component as compared to the horizontal one.
At short times, in the ballistic regime, this result can be interpreted by considering pairs of different orientations at the release time, and is a signature of the anisotropy of Eulerian second-order functions.
Rotation also enhances the distortion of triads and tetrads also present in homogeneous and isotropic turbulence.
In particular, at long times, the flattening of tetrads increases with the rotation rate.
The maximal dimension of triads and tetrads is shown to be preferentially aligned with the rotation axis, in agreement with our observations for pairs.
\end{abstract}

\maketitle

\section{Introduction}

Turbulence has been for a long time investigated by using the Eulerian point of view.
However, the fundamental mechanisms of turbulent flows, as well as their mixing and transport properties, can be more naturally understood in the Lagrangian framework.
In particular, the relative dispersion of a pair of tracers is closely related to the growth of a blob of passive scalar in a turbulent flow.
The Lagrangian approach to turbulence has seen significant developments since the beginning of the century, on the theoretical, experimental and numerical sides~\cite{Yeung2002,Toschi2009}.
Numerous works have been devoted to the investigation of the Lagrangian dynamics of a single fluid particle~\cite{LaPorta2001,Mordant2002,Arneodo2008}.

The study of \emph{pair} dispersion~\cite{Sawford2001} was pioneered almost a century ago by Richardson~\cite{Richardson1926}, who put forward the idea that turbulence strongly accelerates the separation between tracer particles when their separation distance lies in the inertial range of the flow, as they are advected by turbulent eddies in different directions.
This mechanism explains the exceptional capacity of turbulence to mix and diffuse passive scalars.
According to Richardson's classical picture, this explosive separation regime is expressed as $\Lmean{|\delta \xp(t)|^2} = g \diss t^3$, where $\Lmean{|\delta \xp|^2}$ is the mean squared distance between the particles, $\diss$ is the turbulence energy dissipation rate, and $g$ is known as Richardson's constant.
This relation can be predicted from dimensional analysis in the framework of Kolmogorov's K41 theory~\cite{Salazar2009}.
At short times, when the influence of turbulent fluctuations is not important enough to decorrelate the relative motion of the tracers, the mean-square separation $\Lmean{\DeltaFull^2}(t) = \Lmean{|\delta \xp(t)-\delta \xp(0)|^2}$ grows ballistically, with a $\sim t^2$ scaling and an average separation velocity given by the second-order Eulerian velocity structure function~\cite{Batchelor1950}.
At late times, when the dynamics of the two particles become decorrelated, the application of Taylor's law~\cite{Taylor1922}, valid for the diffusion of a single particle, results in a $\sim t$ scaling for their mean square distance.

While pairs provide information on the growth of a scalar cloud, the dynamics of a cluster of $n\ge 3$ particles can give some insight on the geometric structure of turbulence.
A tetrad ($n=4$) is the minimal configuration allowing to define a volume, and thereby to account for the three-dimensionality of the flow.
Numerical works have been devoted to the investigation of the dynamics of tetrads in homogeneous and isotropic turbulence (HIT)~\cite{Chertkov1999,Pumir2000,Biferale2005b,Hackl2011,Pumir2013}.
These objects have been found to become preferentially coplanar, reflecting the tendency of turbulence to ``flatten'' blobs of fluid.
In two-dimensional turbulence, triads were also found experimentally to be preferentially elongated~\cite{Castiglione2001}.

The above-mentioned investigations were carried out in isotropic turbulence.
However, many turbulent flows of natural (ocean, atmosphere, \ldots) or industrial (turbomachinery, wind turbines, \ldots) relevance are subject to solid-body rotation.
Rotation modifies the dynamics and structure of turbulence by imposing a preferential direction, and thereby a strong anisotropy in the flow.
It leads in particular to the formation of large-scale vortical columnar structures aligned with the rotation axis~\cite{Hopfinger1982,Smith1999,Moisy2011,Godeferd2015,Naso2015}.
However, the effect of such features on the dispersion and deformation of Lagrangian objects has received little attention in the literature.
In geophysical flows, rotation is very often coupled to density stratification.
Single-particle and pair dispersions were investigated numerically in decaying turbulence subject to rotation \emph{and} stratification~\cite{Liechtenstein2006}.
The dynamics of $1\le n\le 4$ particles was investigated in purely stratified turbulence~\cite{vanAartrijk2008}.
To the best of our knowledge, the only investigations of purely rotating turbulence carried out in the Lagrangian framework considered a single particle~\cite{Cambon2004,DelCastello2011,DelCastello2011a,Alards2017,Maity2019,Buaria2020a}.

In this work, we investigate the relative dispersion and geometry of clusters of two, three and four tracers in homogeneous rotating turbulence, by using direct numerical simulations at different rotation rates.
We begin in Sec.~\ref{sec:def_num_sim} by recalling the definitions of the size and geometry indicators that will used throughout the paper, and by describing our numerical method.
Section~\ref{sec:flow_char} is then devoted to the characterization of the resulting flows.
The size and geometry statistics obtained for pairs, triads and tetrads are presented and discussed in Sec.~\ref{sec:lagr_results}.
Our conclusions are finally provided in Sec.~\ref{sec:concl}.

\section{Definitions and numerical simulations}\label{sec:def_num_sim}

\subsection{Lagrangian multi-particle clusters}

\subsubsection{Geometric definitions}%
\label{sub:geometric_definitions}

The present work focuses on the time evolution of the size, shape and orientation of clusters of $n$ tracer particles in three-dimensional rotating turbulence, with $n = 2$ (pairs), $3$ (triads) and $4$ (tetrads).
In general, the geometry of such an $n$-particle cluster with positions $(\xp_1, \ldots, \xp_n)$ can be fully described by a set of $n - 1$ reduced separation vectors~\cite{Chertkov1999, Hackl2011} defined by
\begin{equation}
  \label{eq:reduced_separation_vectors}
  \rhovec{m} =
  \sqrt{\frac{m}{m + 1}} \left[ \xp_{m + 1} - \frac{1}{m} \sum_{i = 1}^m \xp_i \right],
  \quad m = 1, \ldots, n - 1.
\end{equation}
Here, the scaling factor is such that, if all positions $\xp_i$ within a cluster are statistically independent, then the variance of each $\rhovec{m}$ matches the variance of the positions $\xp_i$~\cite{Hackl2011}.

\begin{figure}[t]
  \centering
  \includegraphics[width = 0.6\textwidth]{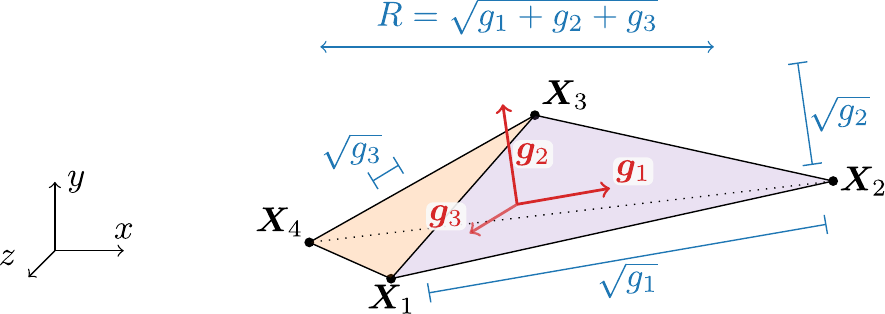}
  \caption{%
    Schematic illustration of a tetrad and its eigendecomposition.
    The tetrad gyration radius $R$ and the square-root eigenvalues $\sqrt{\gval{i}}$ are represented to-scale.
  }%
  \label{fig:diagram_tetrad}
\end{figure}

The geometrical properties of a cluster are then contained in the $3 \times 3$ moment of inertia-like tensor $\InertiaTensorMat$~\cite{Pumir2000} defined by $\InertiaTensor_{ij} = \sum_{k = 1}^{n - 1} \rho_i^{(k)} \rho_j^{(k)}$.
In particular, the squared radius of gyration -- characterizing the linear size of the cluster -- is given by
\begin{equation}
  \label{eq:def_R2}
  R^2 \equiv
  \frac{1}{2n} \sum_{i = 1}^n \sum_{j = 1}^n |\xp[i] - \xp[j]|^2
  = \sum_{k = 1}^{n - 1} {\big| \rhovec{k} \big|}^2
  = \InertiaTensor_{11} + \InertiaTensor_{22} + \InertiaTensor_{33} = \gval{1} + \gval{2} + \gval{3},
\end{equation}
where $\gval{1}$, $\gval{2}$ and $\gval{3}$ are the eigenvalues of $\InertiaTensorMat$.
For $n \le 4$, the cluster volume is given by $V = {(\gval{1} \gval{2} \gval{3})}^{1/2} / 3$~\cite{Hackl2011}.
As in previous works~\cite{Pumir2000, Hackl2011}, we will thereafter sort the eigenvalues in decreasing order, $\gval{1} \ge \gval{2} \ge \gval{3} \ge 0$.
Furthermore, the orientation dynamics of a multi-particle cluster can be characterized by the respective eigenvectors $\gvec{1}$, $\gvec{2}$ and $\gvec{3}$, where $\gvec{1}$ and $\gvec{3}$ respectively represent the directions of maximum and minimum elongation of the cluster.
The eigendecomposition of a four-particle cluster (a tetrad) is illustrated in Fig.~\ref{fig:diagram_tetrad}.

For a regular tetrad with edges of equal length $r_0$, the eigenvalues of $\InertiaTensor$ satisfy $\gval{1} = \gval{2} = \gval{3} = r_0^2$.
In contrast, for a flattened (pancake-like) tetrad, $\gval{1} \sim \gval{2} \gg \gval{3}$; while for a needle-like tetrad $\gval{1} \gg \gval{2}, \gval{3}$~\cite{Pumir2000}.
The shape of a tetrad may also be characterized by the non-dimensional shape parameter $\Lambda = V^{2/3} / R^2$~\cite{Hackl2011}.
For instance, since the volume $V$ of pancake and needle-like tetrads is negligible compared to the gyration radius $R$, one has $\Lambda \approx 0$ in those cases.
Conversely, for a regular tetrad, $R^2 = 3 r_0^2$ and $V = r_0^3 / 3$, which gives the maximum possible value of the shape parameter, $\LambdaRegular \equiv 3^{-5/3} \approx 0.16$.

In the case of triads, the moment of inertia tensor has rank $r \le 2$, and therefore $\gvalTriad{3} = 0$.
The shape of a triad with gyration radius $R$ is commonly quantified by the ratio $w = A / \AreaRegular$ between its area $A$ and the area $\AreaRegular = \sqrt{3} R^2 / 4$ of an equilateral triangle with the same gyration radius~\cite{Pumir1998, Shraiman1998, Castiglione2001, Hackl2011}.
The triad shape parameter $w$ thus takes values between $0$ for a collinear triangle with zero area, and $1$ for an equilateral triangle.
The specification of the triad shape is completed by the Euler angle $\chi = \frac{1}{2} \arctan{\left\{
    2 \left( \rhovec{1} \cdot \rhovec{2} \right) /
    \left( \rhovec{2} \cdot \rhovec{2} - \rhovec{1} \cdot \rhovec{1} \right)
\right\}}$, which quantifies the triangle symmetry~\cite{Shraiman1998, Castiglione2001}.
After taking into account particle relabeling symmetry, values of $\chi$ can be reduced to the interval $[0, \pi / 6]$,
where the two limits respectively represent isosceles triangles whose unequal side is smaller and larger than the equal sides~\cite{Castiglione2001}.
This angle is ill-defined for equilateral triangles.
Alternatively, and similarly to tetrads, the relative importance of the eigenvalues $\gvalTriad{1}$ and $\gvalTriad{2}$ can provide information on the shape of a triad ($\gvalTriad{1} = \gvalTriad{2}$ for equilateral triangles), while the statistics of the respective eigenvectors $\gvecTriad{1}$ and $\gvecTriad{2}$ provide information on the preferential orientation of triads in the flow.

As for particle pairs, it is easy to see from Eqs.~\eqref{eq:reduced_separation_vectors} and \eqref{eq:def_R2} that their gyration radius $R$ is proportional (but not equal) to the particle separation.
Namely, $R = |\delta\xp| / \sqrt{2}$, where $\delta\xp \equiv \xp[2] - \xp[1]$ is the particle separation vector.
For consistency with tetrads and triads, throughout this work we consider the pair gyration radius $R$ (and its $\DeltaFull$ variant defined in Sec.~\ref{sub:ballistic_regime}) instead of the equivalent pair separation $\delta\xp$ that is more commonly used in the relative pair dispersion literature.

\subsubsection{Ballistic separation regime}%
\label{sub:ballistic_regime}

In turbulent flows, the separation of a pair of fluid tracers tagged at an initial time $t = 0$ follows, at short times, a ballistic regime of the form $\delta\xp(t) - \delta\xp(0) \approx \delta \vp_0 \, t + \frac{1}{2} \delta \ap_0 \, t^2$, where $\delta\vp_0$ and $\delta\ap_0$ are the initial relative velocity and acceleration of the particle pair~\cite{Batchelor1950, Bourgoin2006, Ouellette2006a}.
This regime is valid at times over which the influence of turbulent fluctuations is not important enough to decorrelate the relative motion of the tracers, and results from a Taylor expansion of the particle trajectories.
The ballistic regime can be readily generalized to $n$-particle clusters via the reduced separation vectors [Eq.~\eqref{eq:reduced_separation_vectors}], leading to
\begin{align}
  \DeltaFull^2(t)
  &\equiv
  \sum_{k = 1}^{n - 1} {\big| \rhovec{k}(t) - \rhovec{k}(0) \big|}^2
  \label{eq:def_Delta2}
  \\
  &= \sum_{k = 1}^{n - 1} {\big| \vp[k]_0 \big|}^2 \, t^2
  + \sum_{k = 1}^{n - 1} \vp[k]_0 \cdot \ap[k]_0 \, t^3 + O(t^4)
  \quad \text{for } t \ll t_0,
  \label{eq:Delta2_ballistic}
\end{align}
where $\vp[k]_0 = \left. \diff{\rhovec{k}}{t} \right|_{t = 0}$ and $\ap[k]_0 = \left. \diff[2]{\rhovec{k}}{t} \right|_{t = 0}$ and $t_0$ is a ballistic time scale.
Note that the definition of $\DeltaFull^2$ is very similar to that of the gyration radius $R^2$ in Eq.~\eqref{eq:def_R2}.
In fact, $\DeltaFull^2$ is expected to be approximately equal to $R^2$ at sufficiently long times (as $|\rhovec{k}(t)| \gg |\rhovec{k}(0)|$ for large $t$).
However, at short times, only $\DeltaFull^2$ is expected to rigorously display the $t^2$ scaling associated with the ballistic regime, unlike $R^2$ (as shown later in Fig.~\ref{fig:pairs_R2}).
In particular, note that $\DeltaFull^2 = 0$ at $t = 0$, and for this reason we refer to $\DeltaFull$ as the \emph{change of} gyration radius.
These considerations generalize the distinction between the squared separation $|\delta\xp(t)|^2$ and the squared change-of-separation $|\delta\xp(t) - \delta\xp(0)|^2$ in the case of particle pairs~\cite{Ouellette2006a, Polanco2018}.

The time scale $t_0$ over which the ballistic regime takes place may be estimated, for particle pairs, as $t_0 = \Lmean{|\delta\vp_0|^2} / |\Lmean{\delta\vp_0 \cdot \delta\ap_0}|$~\cite{Bitane2012}.
By this definition, $t_0$ depends on the initial pair separation (or initial cluster scale) $r_0$.
In isotropic turbulence, $\Lmean{|\delta\vp_0|^2} = S_2(r_0)$, where $S_2$ is the second-order Eulerian velocity structure function.
If $r_0$ is in the inertial range of scales, K41 theory predicts $S_2(r_0) \propto (\diss r_0)^{2/3}$~\cite{Frisch1995}, where $\diss$ is the mean energy dissipation rate per unit mass.
Additionally, for $r_0$ in the inertial range, $\Lmean{\delta\vp_0 \cdot \delta\ap_0} \approx -2\diss$~\cite{Falkovich2001, Ott2000, Hill2006}, leading to $t_0 \propto (r_0^2 / \diss)^{1/3}$.
Hence, in this case, $t_0$ is proportional to the turbulent eddy-turnover time at scale $r_0$~\cite{Frisch1995}, as originally suggested by Batchelor~\cite{Batchelor1950}.

\subsection{Numerical simulations}

\subsubsection{Navier-Stokes equations in rotating frame}

We consider a homogeneously rotating turbulent flow described by the incompressible Navier-Stokes equations.
In the rotating reference frame, these read
\begin{gather}
  \diffp{\uvec}{t} + (\uvec \cdot \gradient) \uvec
  = -\frac{1}{\rho} \gradient p + \nu \laplacian \uvec
  + 2\uvec \cross \RotationVec + \Fvec,
  \label{eq:NS_rotating}
  \\
  \divergence \uvec = 0,
  \label{eq:NS_incompressiblity}
\end{gather}
where $\uvec(\xvec, t)$ and $p(\xvec, t)$ are respectively the velocity and the pressure fields in the rotating frame, $\nu$ and $\rho$ are respectively the fluid kinematic viscosity and density, and $\RotationVec$ is the rotation vector.
Note that the Coriolis term $2\uvec \cross \RotationVec$ performs no work, and as such, it has no influence on the global energy balance of the system.
In Eq.~\eqref{eq:NS_rotating}, the pressure field includes the contribution of the centrifugal force $\bm{f}_\text{c} = -\gradient \left|\RotationVec \cross \xvec\right|^2 / 2$.
Throughout this paper, without loss of generality, we take the rotation vector as pointing along the third Cartesian direction, i.e.\ $\RotationVec = \OmegaRot \evec_z = (0, 0, \OmegaRot)$.
We also refer to $\evec_z$ as the \emph{vertical} direction, while directions perpendicular to the rotation axis are \emph{horizontal}.
Besides, $\Fvec(\xvec, t)$ is a forcing term representing an
external energy injection mechanism.
Here, its influence is limited to the large scales of the system, specifically to those Fourier coefficients $\hat{\uvec}(\kvec, t)$ of the velocity field such that $|\kvec| \le \kforcing$, with $\kforcing = 1.5$.
As in previous works~\cite{Pumir1994, Naso2012, Vallefuoco2018}, such modes here obey the truncated Euler equations~\cite{Cichowlas2005} within the sphere $|\kvec| \le \kforcing$ in the rotating frame, while modes $|\kvec| > \kforcing$ obey the incompressible Navier-Stokes equations \eqref{eq:NS_rotating} and \eqref{eq:NS_incompressiblity}.

\begin{table*}[tb]
  \centering
  \caption{%
    Physical and numerical simulation parameters.
    $\RossbyL = \urms / (2 L \OmegaRot)$, large-scale Rossby number;
    $\RossbyOmega = \vortrms / (2 \OmegaRot)$, small-scale Rossby number;
    $\ReLambda = \urms \lambdaTaylor / \nu$, Reynolds number based on the Taylor microscale $\lambda$;
    $L = (\pi / \urms^2) \int_0^{\infty} E(k) / k \, dk$, Eulerian integral length scale;
    $\eta = (\nu^3 / \diss)^{1/4}$, Kolmogorov length scale;
    $\TL^i = \int_0^\infty \corrvel_i(\tau) \, d\tau$, horizontal and vertical Lagrangian integral time scales;
    $\taueta = \sqrt{\nu / \diss}$, Kolmogorov time scale;
    $\Lmean{u_i^2}^{1/2} / u_\eta$, fluctuations of the horizontal and vertical components of Eulerian velocity
    normalized by $u_\eta = \eta / \tau_\eta$;
    $k_\text{max} = N / 3$, maximum resolved Fourier wave number in the simulations.
  }
  \label{tab:simulation_parameters}
  \begin{ruledtabular}  
  \begin{tabular}{cccccccccc}
    Run & $\RossbyL$  & $\RossbyOmega$ & $\ReLambda$ & $L / \eta$ & $\TL^x / \taueta$ & $\TL^z / \taueta$ & $\Lmean{u_x^2}^{1/2} / u_\eta$ & $\Lmean{u_z^2}^{1/2} / u_\eta$ & $k_\text{max} \eta$
        \\
    \midrule
    1 & $\infty$       & $\infty$   & \num{107} & \num{74.8} & \num{10.8} & \num{10.1} & \num{5.22} & \num{5.13} & \num{1.49} \\
    2 & \num{\RoLarge} & \num{1.29} & \num{115} & \num{73.4} & \num{5.3}  & \num{10.4} & \num{5.29} & \num{5.55} & \num{1.66} \\
    3 & \num{\RoMid}   & \num{0.52} & \num{138} & \num{69.4} & \num{3.6}  & \num{11.4} & \num{5.76} & \num{6.12} & \num{1.85} \\
    4 & \num{\RoSmall} & \num{0.18} & \num{162} & \num{64.5} & \num{4.2}  & \num{15.0} & \num{5.99} & \num{6.98} & \num{2.19} \\
  \end{tabular}
  \end{ruledtabular}
\end{table*}

The above equations are numerically solved in a three-dimensional periodic domain of size $(2\pi)^3$ using direct numerical simulations (DNS).
The solver uses a standard Fourier pseudo-spectral method, including the 2/3 dealiasing rule to suppress discretization errors stemming from the non-linear term of Eq.~\eqref{eq:NS_rotating}.
The temporal advancement is performed using an explicit third-order Adams-Bashforth scheme for the non-linear term, while the viscous term is treated exactly using an integrating factor technique.
The simulations reported here are performed on $N^3 = 256^3$ collocation points for different rotation rates $\OmegaRot$.
The rotation rate can be non-dimensionalized in terms of a Rossby number, defined as the ratio between timescales associated with rotation and with turbulent fluctuations.
We define here large-scale and small-scale Rossby numbers, respectively as $\RossbyL = \urms / (2L \OmegaRot)$ and $\RossbyOmega = \vortrms / (2 \OmegaRot)$,
where $\urms = \sqrt{\Emean{\uvec^2} / 3}$ and $\vortrms = \sqrt{\Emean{\vortvec^2} / 3}$ are respectively the velocity and vorticity standard deviations.
Here, $L = (\pi / \urms^2) \int_0^{\infty} E(k) / k \, dk$ is the integral length scale of the flow, with $E(k)$ being the kinetic energy spectrum.
The Rossby numbers for the different runs considered here are listed in Table~\ref{tab:simulation_parameters}.
Note that we also consider a reference run without rotation (Run 1), corresponding to homogeneous isotropic turbulence.
Also listed in Table~\ref{tab:simulation_parameters} are the Reynolds numbers $\ReLambda = \lambdaTaylor \urms / \nu$ associated to each run -- where $\lambdaTaylor = \sqrt{5 \Emean{\uvec^2} / \Emean{\vortvec^2}}$ is the Taylor microscale~\cite{Tennekes1972} -- and the maximal resolved Fourier wave number $k_\text{max}$.
The degree of scale separation in the Eulerian frame may also be characterized by the ratio, reported in Table~\ref{tab:simulation_parameters}, between the integral length scale $L$ and the Kolmogorov scale $\eta$.
Finally, we give in Table~\ref{tab:simulation_parameters} the values of $\TL^i$, the Lagrangian integral time scales in the horizontal and vertical directions (see Sec.~\ref{sec:autocorr}), as well as the standard deviations of the Eulerian velocity components in the same directions, $\Lmean{u_i^2}^{1/2}$.

\subsubsection{Lagrangian particle tracking}%
\label{sub:lagrangian_tracking}

To characterize the flow in the Lagrangian framework, the tracking of fluid particles is performed within the Navier-Stokes solver.
Particles are considered as ideal Lagrangian tracers obeying the equation of motion $\diff{\xp}{t}(t) = \uvec[\xp(t), t]$, where $\xp(t)$ is the instantaneous tracer position.
The tracer equation of motion is discretized in time using the same Adams-Bashforth scheme as for the Eulerian fields.
Instantaneous fluid velocities at particle positions are obtained using a sixth-order Lagrange interpolation scheme~\cite{Naso2012}.

To investigate the dispersion of multiparticle Lagrangian clusters, we track the motion of particle tetrads initialized as the vertices of regular tetrahedrons of different sizes.
Particles are released at a time $t = 0$ at which the Eulerian fields have already achieved a statistically-steady state.
The initial cluster size is denoted $r_0$, and is defined such that the initial gyration radius of an $n$-particle cluster is $R^2_0 = (n - 1) r_0^2$.
As a result of this convention, the initial separation between any pair of particles is $r_0 \sqrt{2}$.
The values $r_0 / \eta = 1/2$, $1$, $4$, $16$ and $64$ are considered, where $\eta = (\nu^3 / \diss)^{1/4}$ is the Kolmogorov scale of the flow.
Tetrads are initialized at random positions throughout the computational domain.
For each considered rotation rate, a total of 249\,856 particles (81\,920 tetrads) is initialized at two independent time instants.
The tetrads' initial orientations are also random, i.e.\ there is initially no preferential alignment of the eigenvectors $\gvec{i}$ with the Cartesian axes $\evec[j]$.
Note that the faces and edges of each regular tetrad respectively define 4 equilateral triangles and 6 Lagrangian pairs, enabling in addition the study of two- and three-particle clusters.

Furthermore, to investigate the role of flow anisotropy on relative dispersion, we consider a second initialization scheme also considered in previous works~\cite{Hackl2011, vanAartrijk2008, Polanco2018}.
Tetrads are then set-up as trirectangular tetrahedrons, so that three of the edges are mutually-orthogonal and aligned with the three Cartesian axes, and have equal lengths $r_0 \sqrt{2}$.
In this case, a total of 131\,072 particles (40\,960 tetrads) is tracked for each rotation rate.
This scheme will be used for the tracking of particle pairs whose initial relative orientation is either parallel or perpendicular to the rotation axis.
Concretely, this second dataset is used in
Sec.~\ref{par:effect_of_orientation} (Fig.~\ref{fig:pairs_R2_oriented}) to
investigate such orientation effects.

\section{Flow characterization}\label{sec:flow_char}

\subsection{Eulerian flow features}

\begin{figure*}[tb]
  \centering
  \includegraphics[width = 0.9\textwidth]{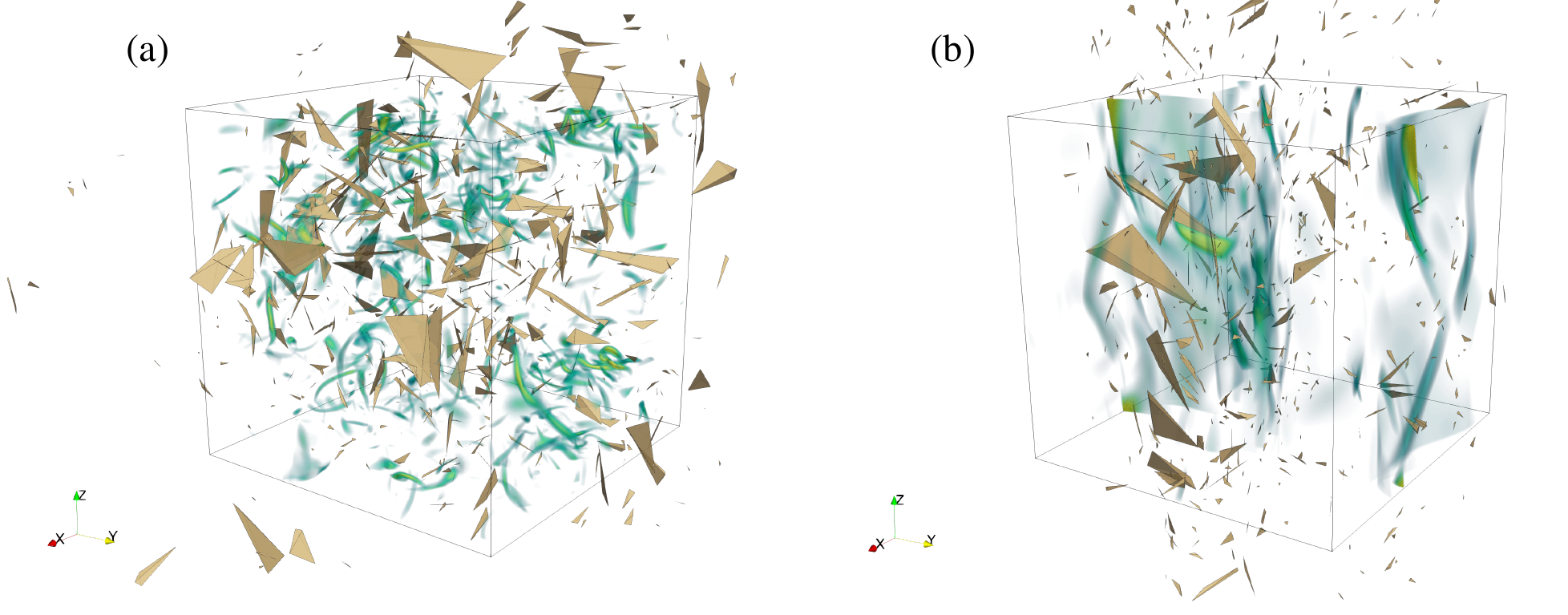}
  \caption{%
    Visualization of Lagrangian tetrads in turbulent flow from DNS at resolution $N^3 = 256^3$.
    Green colors represent high values of the local vorticity magnitude $|\vortvec(\xvec)|$.
    The periodic computational domain is represented by thin gray lines.
    (a)~Isotropic case (Run 1, $\RossbyL = \infty$).
    (b)~Strongly-rotating case (Run 4, $\RossbyL = \RoSmall$).
    The rotation axis is aligned with the $z$ direction.
    Snapshots are taken at time delays $t / \taueta = $ 15 (a) and 12 (b) following the release of the particles.
  }%
  \label{fig:visualisation}
\end{figure*}

High-vorticity regions are visualized in Fig.~\ref{fig:visualisation} for a non-rotating and a highly-rotating flow.
In the highly-rotating case [Fig.~\ref{fig:visualisation}(b)], the presence of large-scale vortex columns (volume-rendered in green) aligned with the axis of rotation is clear, contrasting with the isotropic vorticity distribution in the absence of rotation [Fig.~\ref{fig:visualisation}(a)].
Such vortex columns are a well-known feature of rotating turbulence, which have been observed in a number of experiments and numerical simulations (e.g.~\cite{Hopfinger1982,Smith1999,Moisy2011,Godeferd2015}).

\begin{figure*}[tb]
  \centering
  \includegraphics[width = \textwidth]{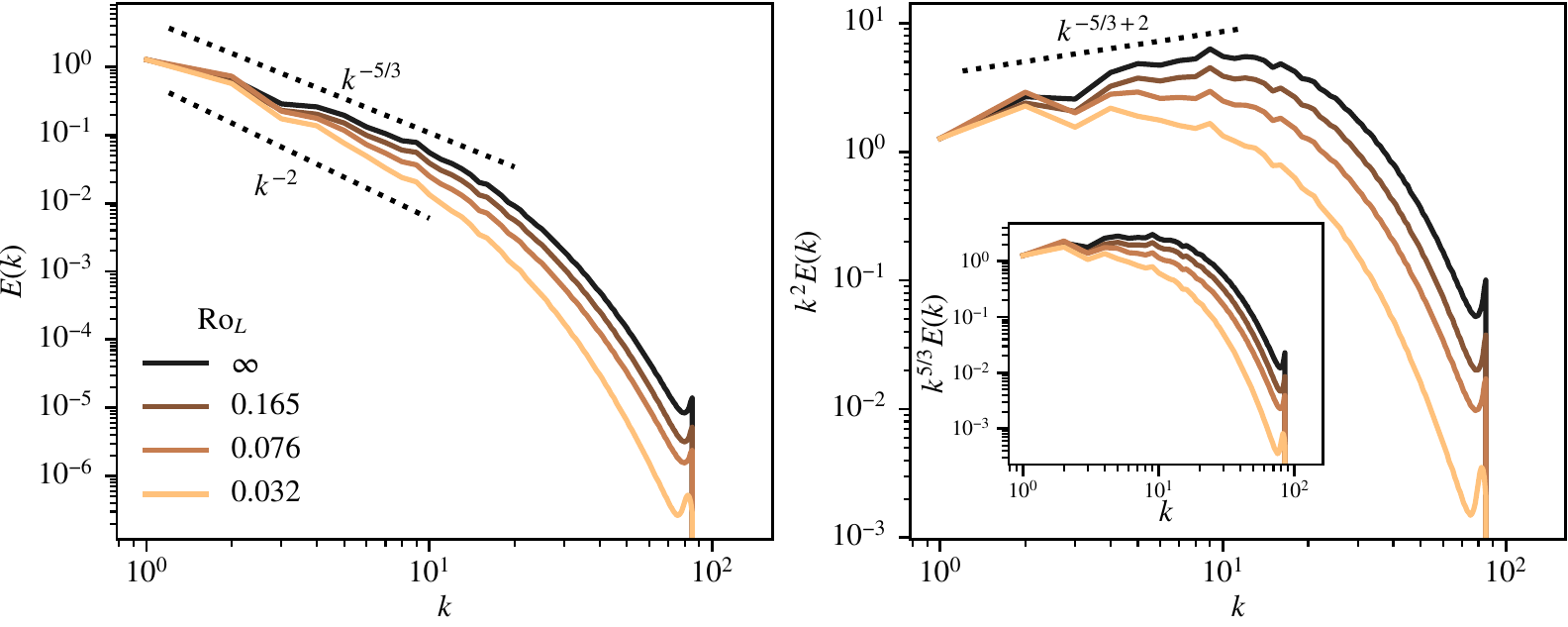}
  \caption{%
    Kinetic energy spectra for different rotation rates.
    (a)~Uncompensated spectra.
    (b)~Spectra compensated by $k^2$.
    Inset: spectra compensated by $k^{5/3}$.
  }%
  \label{fig:spectra}
\end{figure*}

To complement the above qualitative description of the different flows, we plot in Fig.~\ref{fig:spectra} the kinetic energy spectra obtained in the different runs considered here.
As the rotation rate is increased, the energy spectrum in the inertial range of scales departs from the $E(k) \sim k^{-5/3}$ power law predicted by Kolmogorov's K41 theory for isotropic turbulence~\cite{Frisch1995} and becomes slightly steeper.
In particular, at the largest rotation rates, the inertial-range spectrum is compatible with the $E(k) \sim k^{-2}$ spectrum previously observed in rapidly-rotating turbulence experiments and simulations~\cite{Baroud2002, Mininni2010, Biferale2016} and predicted using phenomenological arguments~\cite{Zhou1995}.
Due to flow anisotropy, note that the spectral energy content is fully
described by the 2D energy spectrum $\tilde{E}(k_\parallel,
k_\perp)$, where $k_\parallel$ and $k_\perp$ are the
components of the wave number vector $\kvec$ respectively parallel and perpendicular to
the rotation axis, which we do not consider here.
The reader is referred to previous works~\cite{Smith1999, Thiele2009,
Mininni2010, Godeferd2015} for more details on directional spectra in
homogeneous rotating turbulence.

Also shown in Fig.~\ref{fig:visualisation} are the instantaneous positions of a subset of Lagrangian tetrads in each flow, at a time $t / \taueta \sim 10$ after which they have been substantially dispersed by the flow.
In both cases, most tetrads appear as strongly-flattened structures, far from their initial regular shape.
Moreover, in the rotating case [Fig.~\ref{fig:visualisation}(b)], tetrads appear to be preferentially stretched along the rotation axis -- and have traveled longer distances in the vertical direction compared to the horizontal ones -- while they are mostly flattened in the horizontal directions.

\subsection{Single-particle velocity autocorrelation and integral timescales}
\label{sec:autocorr}

The velocity autocorrelation along fluid paths quantifies the loss of memory of the velocity of a fluid particle over time.
It provides important information on the Lagrangian structure of turbulent flows, including the timescales relevant to dispersion phenomena, as well as characterizing the anisotropy of such flows.
The Lagrangian velocity autocorrelation is defined as
\begin{equation}
  \corrvel_{i}(t) = \frac{\Lmean{v_i(t_0) \, v_i(t_0 + t)}}{\Lmean{v_i^2}},
  \quad
  i \in \{x, y, z\},
\end{equation}
where $\vp$ is the velocity of a particle.
Averages are performed over individual Lagrangian tracers and initial times $t_0$, and repeated indices do not imply summation.
Note that the above expression assumes a statistically-homogeneous and steady turbulent flow ($\corrvel_i$ does not depend on $t_0$ nor on the initial location of the particle), which is the case studied here.

\begin{figure}[tb]
  \centering
  \includegraphics[width = \textwidth]{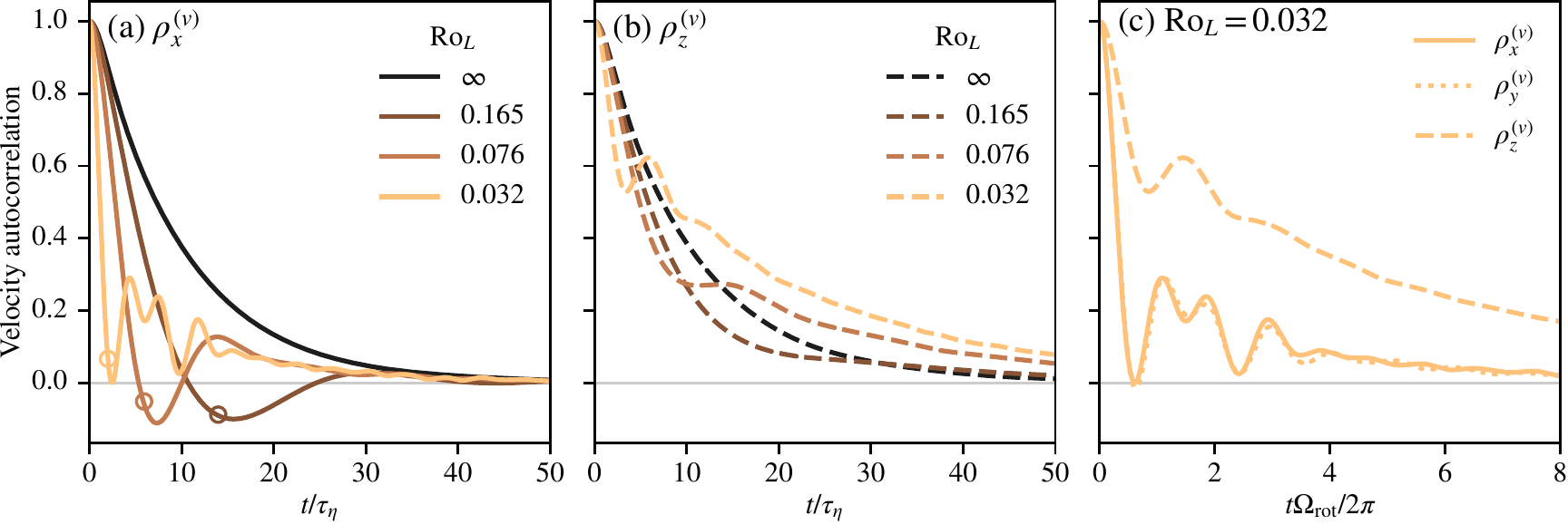}
  \caption{%
    Lagrangian velocity autocorrelations.
    (a)~Horizontal component $\corrvel_x$ (orthogonal to the rotation axis) for different rotation rates.
        Circles mark the time $t = \Trot / 2$ after half a rotation period, where $\Trot = 2\pi / \OmegaRot$.
    (b)~Vertical component $\corrvel_z$ (parallel to the rotation axis) for different rotation rates.
    (c)~All components for $\RossbyL = \RoSmall$ (Run 4).
    Time is normalized (a-b) by the Kolmogorov timescale $\taueta$ and (c) by the rotation period $\Trot$.
  }%
  \label{fig:velocity_correlations}
\end{figure}

In Fig.~\ref{fig:velocity_correlations}(a), the Lagrangian autocorrelation of the horizontal velocity component $v_x$ is plotted for different rotation rates.
In the isotropic case (infinite $\RossbyL$), $\corrvel_x$ decays exponentially, with a long time behavior $\sim e^{-t/T_L}$ allowing to define the integral time $T_L$, as expected~\cite{Yeung2002}.
As the rotation rate is increased (decreasing $\RossbyL$), the horizontal velocity decorrelates faster relative to the dissipation timescale $\taueta$.
At intermediate rotation rates, the autocorrelation becomes negative before reaching zero at long times, in contrast to the monotonically-decreasing correlation in the absence of rotation.
This may be explained by the effect of large-scale vortex columns in rotating turbulence, which are aligned with the rotation axis.
These are expected to induce helical trajectories on Lagrangian particles, leading to periodic oscillations of the horizontal velocity components.
As shown by the circles in Fig.~\ref{fig:velocity_correlations}(a), the fluctuations of the autocorrelation curves have a period close to, but slightly larger than the rotation period $\Trot = 2\pi / \OmegaRot$.
Note that the same behavior was displayed by the Lagrangian autocorrelation of horizontal velocity in recent simulations of rotating \emph{and} stratified turbulence~\cite{Buaria2020a}.
Furthermore, in all cases, the correlations reach zero at approximately the same time $t / \taueta \approx 35$.
On a side note, we have checked that the autocorrelation of $v_y$ is statistically identical to that of $v_x$, as illustrated in Fig.~\ref{fig:velocity_correlations}(c) for the highest rotation rate.

At high rotation rates ($\RossbyL = \RoSmall$), the autocorrelation stays positive and displays intense fluctuations.
We have verified the robustness of this observation using an independent particle dataset (not shown here).
The fact that the correlation does not become negative in this case is
somewhat at odds with the above picture of columnar vortices inducing helical
Lagrangian trajectories.
This shows that the complexity of the Lagrangian particle dynamics goes far
beyond this very simple picture.
To summarize, for the three finite Rossby numbers considered, $\corrvel_x(t)$ first decays in time, and starts increasing shortly after $t = \Trot/2$ (see circles in the figure). The initial decay is increasingly faster at increasing rotation rate: at $t = \Trot/2$, $\corrvel_x(t)$ is negative for the two largest Rossby numbers, but for $\RossbyL = 0.032$ it is still positive.

In the presence of rotation, the vertical velocity autocorrelation $\corrvel_z$ [Fig.~\ref{fig:velocity_correlations}(b)] displays a behavior which differs from that of the horizontal component.
This is a clear signature of the anisotropy induced by rotation.
The most striking difference is that $\corrvel_z$ displays very weak oscillations, even at the highest rotation rate considered.
Moreover, the vertical velocity $v_z$ stays correlated for a time longer than the horizontal components.
At long times, the decay rate of $\corrvel_z$ is slower than in isotropic turbulence, an effect that becomes more pronounced at higher rotation rates.
These features reflect the quasi-two-dimensionalization of turbulence by rotation, according to which horizontal and vertical motions decouple and the velocity field becomes nearly invariant along the vertical direction.

The effect of rotation on the autocorrelation of the vertical velocity reported here [Fig.~\ref{fig:velocity_correlations}(b)] is in agreement with experimental measurements available in the literature~\cite{DelCastello2011a}.
As for the horizontal velocity, these experiments display a trend opposite to the present one, i.e., rotation was shown to increase the correlation time of the horizontal velocity components.
The precise reason of this disagreement is not clear, but it is worth mentioning that the numerical and experimental flows are certainly very different, the latter displaying strong confinement effects in the vertical direction, as evidenced by the distribution of vertical velocity in the cited experiments.

The time integration of the autocorrelation functions displayed in Fig.~\ref{fig:velocity_correlations} allows us to calculate horizontal and vertical Lagrangian timescales, $T_L^x$ and $T_L^z$.
In agreement with the features of $\corrvel_x$ and $\corrvel_z$, the values provided in Table~\ref{tab:simulation_parameters} show that the vertical (resp.\ horizontal) integral timescale increases (resp.\ decreases) at decreasing Rossby number, i.e.\ at increasing rotation rate.
The interpretation of $T_L^z$ is clearer than that of $T_L^x$ because of the monotonous behavior of $\corrvel_z$.
For this reason we will thereafter use it to characterize the large Lagrangian timescale of the flow.

\section{Multi-particle statistics} \label{sec:lagr_results}

We discuss now the dispersion of multi-particle Lagrangian clusters in homogeneous rotating turbulence.
We start in Sec.~\ref{sub:dispersion_pairs} by looking at the relative dispersion of particle pairs.
Then, in Sec.~\ref{sub:dispersion_tetrads}, we study the time evolution of the size, shape and orientation of Lagrangian tetrads.
We finish, in Sec.~\ref{sub:dispersion_triads}, with a description of the dynamics of Lagrangian triangles in rotating flows.

\subsection{Pair dispersion}%
\label{sub:dispersion_pairs}

We consider in this section the time evolution of the relative separation between pairs of tracer particles.
This is the most basic Lagrangian observable providing information on the growth of a fluid patch (or the spreading of an ideal passive scalar) in a turbulent flow.
We investigate the evolution of the total mean-squared separation between particle pairs, followed by a detailed characterization of the anisotropy of pair dispersion when the flow is subject to global rotation.

\subsubsection{Total mean-squared separation}

\begin{figure}[tb]
  \centering
  \includegraphics[width = \textwidth]{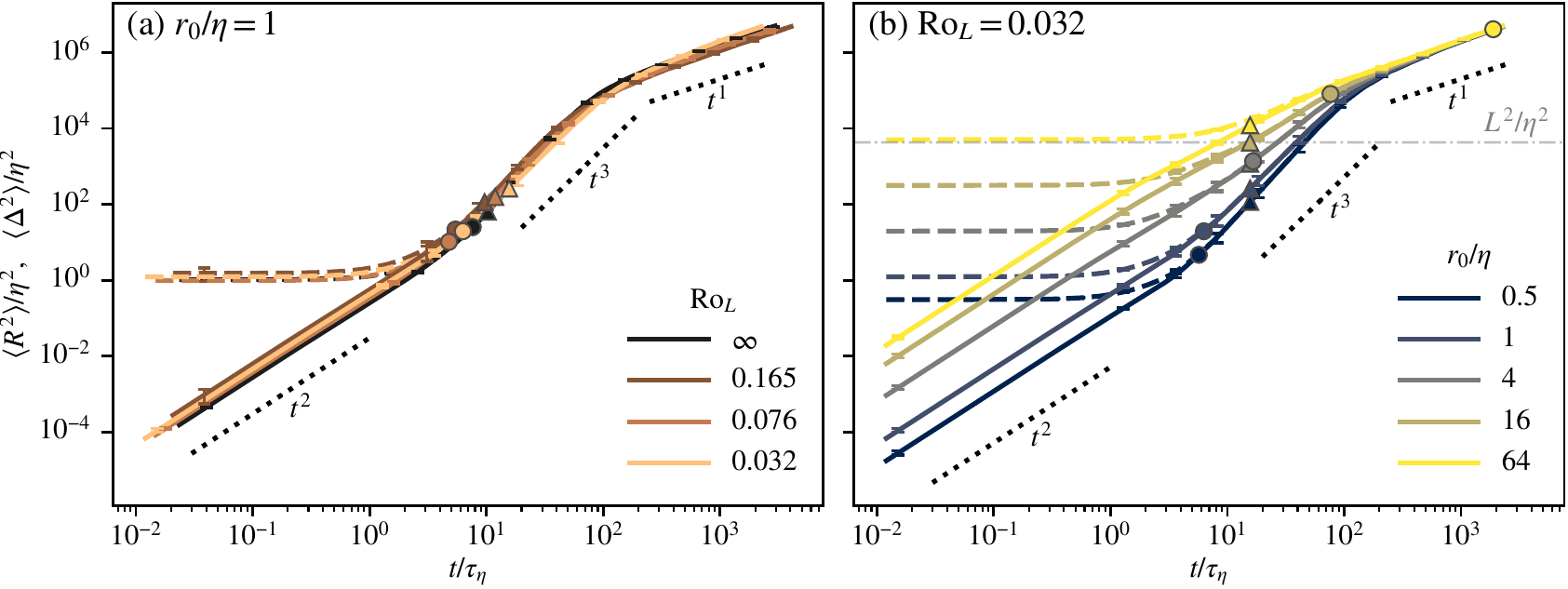}
  \caption{%
    Mean-squared separation between Lagrangian pairs.
    Dashed lines, mean-squared gyration radius $\Lmean{R^2}$ [Eq.~\eqref{eq:def_R2}];
    solid lines, mean-squared \emph{change} of gyration radius $\Lmean{\DeltaFull^2}$ [Eq.~\eqref{eq:def_Delta2}].
    Distances are normalized by the Kolmogorov scale $\eta$.
    (a)~Pairs of initial scale $r_0 = \eta$ and different rotation rates.
    (b)~Pairs of different initial scales $r_0$ under intense rotation ($\RossbyL = \RoSmall$, Run 4).
    Markers represent the ballistic time scale $t_0 = \Lmean{(\delta\vp_0)^2} / |\Lmean{\delta\vp_0 \cdot \delta\ap_0}|$ (circles) and the vertical Lagrangian integral time scale $T_L^z$ (triangles) for the different cases.
    Note that, for particle pairs, the gyration radius $R$ is directly related to the pair separation vector $\delta\xp$ as $R(t)^2 = |\delta\xp(t)|^2 / 2$.
    Equivalently, $\DeltaFull(t)^2 = |\delta\xp(t) - \delta\xp(0)|^2 / 2$.
    In (b), the horizontal dash-dotted line corresponds to $L^2 / \eta^2$, where $L$ is the Eulerian integral length scale.
    Error bars are obtained after splitting each dataset onto 4 disjoint sub-datasets.
    The size of the error bars corresponds to the standard deviation of the statistics computed from each separate sub-dataset.
    Throughout our results, error bars are rather small, especially when plotted in logarithmic scale.
    In the following, for ease of readability, error bars are not shown for quantities plotted in logarithmic scale.
  }%
  \label{fig:pairs_R2}
\end{figure}

We first consider the mean-squared gyration radius $\Lmean{R^2}$ associated to particle pairs.
We recall that the gyration radius $R$ is, up to a factor $\sqrt{2}$, equal to the pair separation $|\delta\xp|$.
The mean-squared gyration radius is represented in Fig.~\ref{fig:pairs_R2} (dashed lines) as a function of time, for different rotation rates $\OmegaRot$ and for different initial scales $r_0$.
As seen in the figure, the mean-squared separation stays approximately constant over a time $t \approx \tau_\eta$, before reaching different regimes where the average particle separation rate clearly speeds-up.
Before discussing the different separation regimes at $t>\tau_\eta$, let us first note that the short-time dynamics may be better described by the mean-squared \emph{change} of separation $\Lmean{\DeltaFull^2}$~\cite{Salazar2009}, defined in Eq.~\eqref{eq:def_Delta2} and represented in Fig.~\ref{fig:pairs_R2} by solid lines.
Indeed, this quantity neatly displays the ballistic separation regime $\Lmean{\DeltaFull^2} \sim t^2$ predicted by Eq.~\eqref{eq:Delta2_ballistic} at times $t \lesssim \tau_\eta$.
As suggested by Eq.~\eqref{eq:Delta2_ballistic} and confirmed by previous works~\cite{Bitane2012, Polanco2018}, the duration of this regime is controlled by the ballistic time scale $t_0$, represented by circles in Fig.~\ref{fig:pairs_R2}.
For $t \gtrsim t_0$, the curves for $\Lmean{R^2}$ and $\Lmean{\DeltaFull^2}$ are indistinguishable, as the instantaneous separation $|\delta\xp|$ becomes much larger than the initial separation $|\delta\xp_0|$.

At long times, all cases asymptotically display the diffusive regime $\Lmean{R^2} \sim t$ first suggested by Taylor~\cite{Taylor1922}.
This regime is achieved when the distance between the particles in a pair is sufficiently larger than the integral length of the flow, such that their motion is completely decorrelated.
As seen in Fig.~\ref{fig:pairs_R2}(b), this regime (and the time at which the transition occurs) is nearly independent of the initial scale $r_0$, and is effectively reached when $R$ is sufficiently larger than $L$.

At intermediate times, different power laws $\Lmean{R^2} \sim t^\alpha$ may be inferred.
Indeed, as shown in Fig.~\ref{fig:pairs_R2}(b), the $\alpha$ exponent displays a strong dependence on the initial scale $r_0$.
In all cases, $t_0$ is close to (and sometimes larger than) the vertical Lagrangian integral time scale $T_L^z$ (represented by triangles), and the scale separation requirement $t_0 \ll t \ll T_L$ for Richardson's regime is therefore never met.
Nevertheless, our results are consistent with recent observations in isotropic turbulence according to which, at finite Reynolds numbers, the regime $\alpha = 3$ traditionally associated with Richardson's law is most likely to be observed for an initial separation $r_0 / \eta \approx 4$~\cite{Elsinga2022, Tan2022}.
We finally stress that, while only the case $\RossbyL = \RoSmall$ is shown, the results are qualitatively similar in the other runs (including the isotropic case).


Up to now, we have not discussed the effect of rotation.
Figure~\ref{fig:pairs_R2}(a) suggests that the impact of rotation on the total relative dispersion is weak.
The most important differences with isotropic turbulence are observed at the highest rotation rate ($\RossbyL = \RoSmall$), in which case the intermediate regime displays a faster separation rate
than in the other runs.
We explain this result in the next subsection, by distinguishing the pair dispersion in the horizontal and in the vertical directions, and by investigating the effect of the initial particle pair orientation.

\subsubsection{Anisotropy of relative dispersion}

\paragraph{Horizontal and vertical dispersion.}

\begin{figure}[tb]
  \centering
  \includegraphics[width = \textwidth]{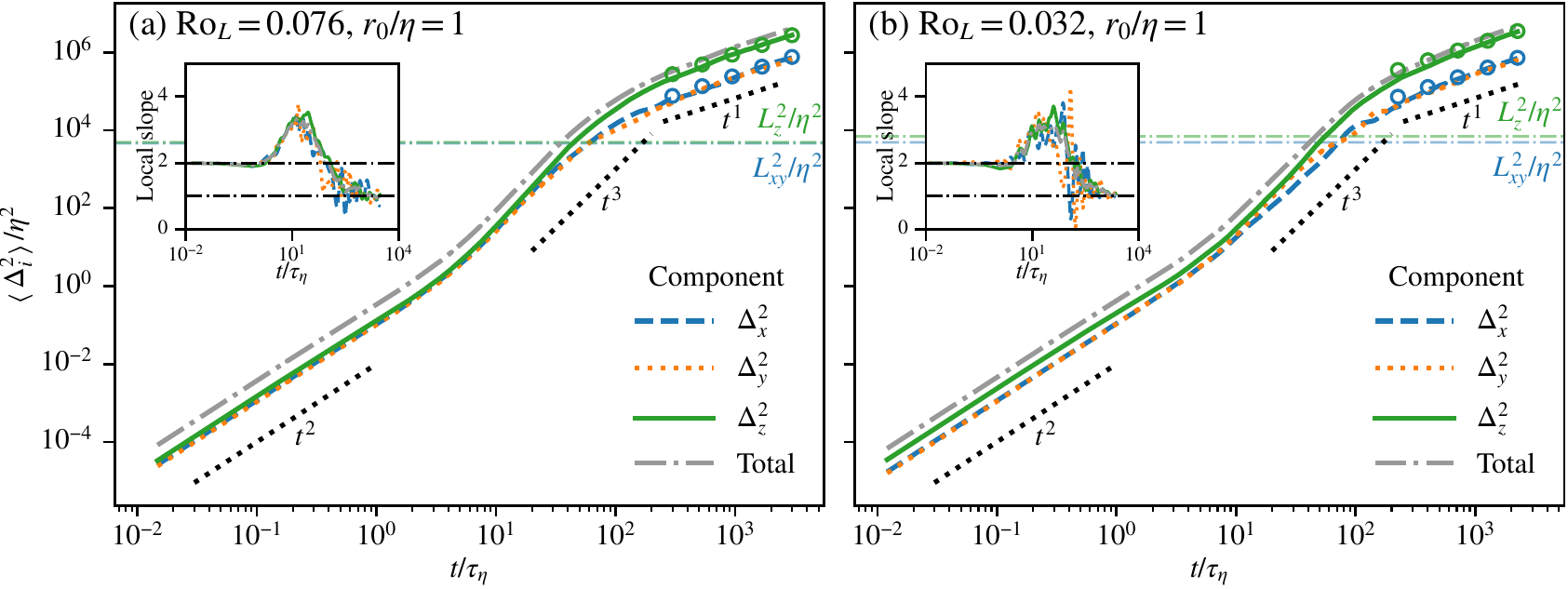}
  \caption{%
    Time evolution of mean-squared separation between Lagrangian pairs along different directions.
    Dashed, dotted and solid lines (in color) represent the squared separation $\Delta^2_i$ along each direction ($i = x, y, z$).
    Dash-dotted gray lines represent the total squared separation $\DeltaFull^2 = \sum_i \Delta^2_i$.
    In all cases, the initial scale is $r_0 = \eta$.
    (a)~Moderately-rotating case (Run 3, $\RossbyL = \RoMid$).
    (b)~Strongly-rotating case (Run 4, $\RossbyL = \RoSmall$).
    Horizontal dash-dotted lines correspond to the Eulerian integral length scales $L_{xy}$ and $L_z$ respectively associated to the horizontal and vertical directions.
    Circles show the diffusive regime predicted by Taylor's theory, $\Delta_i^2(t) = 2 \Lmean{u_i^2} T_L^i t$ for $t \gg T_L$.
    Insets show the local slope of each curve -- obtained as $\alpha = \dd \log \Lmean{\Delta_i^{2}} / \dd \log t$ -- where horizontal lines mark the slopes $\alpha = 1$ and $2$.
  }%
  \label{fig:pairs_R2_components}
\end{figure}

To further elucidate the effect of rotation on
relative dispersion,
we now consider the change of gyration radius along each of the three Cartesian directions, $\Delta_i(t) = |\delta X_i(t) - \delta X_i(0)| / \sqrt{2}$.
Recall that the total change of gyration radius, defined in Sec.~\ref{sub:ballistic_regime}, is given by $\DeltaFull^2 = \sum_i \Delta_i^2$.
In Fig.~\ref{fig:pairs_R2_components}, we plot $\Lmean{\Delta_i^2}$ in cases of moderate and strong rotation, for particle pairs with an initial scale $r_0 = \eta$.
We have verified that the results discussed in the following are also observed for the other considered values of $r_0$.
The local slope of each curve is displayed in the inset, confirming that all
the $\Delta_i$ components scale as $\sim t^2$ at short times (ballistic regime)
and as $\sim t$ at long time (Taylor diffusion regime).

The strongly-rotating case [Fig.~\ref{fig:pairs_R2_components}(b)] displays anisotropy effects at all times, with separation being always more intense in the vertical direction.
In the initial ballistic regime, this can be explained by the vertical velocity $u_z$ decorrelating over shorter distances than the horizontal velocity components.
Indeed, in analogy with Eq.~\eqref{eq:Delta2_ballistic}, the ballistic regime for a given component
can be written as $2 \Lmean{\Delta_i^2}(t) = \Lmean{|\delta\vp_i(0)|^2} \, t^2 + \mathcal{O}(t^3) = S_{2,i}(r_0) \, t^2 + \mathcal{O}(t^3)$,
where $S_{2,i}(r) = \Emean{[u_i(\xvec + \rvec) - u_i(\xvec)]^2}_r$ is the Eulerian second-order structure function associated to the velocity component $u_i$, and spherically-averaged over all increments $|\rvec| = r$.
Therefore, at short times, $\Lmean{\Delta_z^2} > \Lmean{\Delta_x^2}$ is equivalent to $S_{2,z}(r_0) > S_{2,x}(r_0)$, i.e.\ the vertical velocity presents larger spatial fluctuations than the horizontal velocity components.
Such a result will be interpreted further below by investigating the effect of the initial pair orientation or, equivalently, of the increment orientation.
It is important to emphasize the fact that the initial pair separation rates
provide a direct estimation of the Eulerian velocity structure functions at
the scale at which the pairs are initially separated.
This is precisely what is used above and in the following to discuss the
behavior of Eulerian structure functions in homogeneous rotating turbulence.

In the long time limit, the relative dispersion obeys Taylor's regime in all directions.
Concretely, in homogeneous and isotropic turbulence, the mean squared separation is expected to follow $\Lmean{\Delta_i^2} = 2 \Lmean{u_i^2} T_L^i t$ for $t \gg T_L^i$~\cite{Taylor1922, Tennekes1972}.
This prediction is satisfied by our numerical data at all Rossby numbers, as shown by the circles in Fig.~\ref{fig:pairs_R2_components}.
Table~\ref{tab:simulation_parameters} shows that, in the presence of rotation, $\Lmean{u_z^2}$ and $T_L^z$ are respectively larger than $\Lmean{u_x^2}$ and $T_L^x$.
This explains our observation that, at long times, the relative vertical dispersion $\Delta_z$ is faster than the horizontal one, $\Delta_x$ and $\Delta_y$.

In contrast, in the moderately-rotating case [Fig.~\ref{fig:pairs_R2_components}(a)], pairs separate at nearly equal rates along the three directions at short times ($t \lesssim 20\tau_\eta$).
The effects of anisotropy are only evident at longer times, where the separation $\Delta_z$ along the rotation axis becomes larger than the separation in the directions orthogonal to the rotation for the reasons given above.
The anisotropy effect is therefore visible only at large scales for a moderate $\RossbyL = \RoMid$ [Fig.~\ref{fig:pairs_R2_components}(a)], whereas all scales are affected by rotation for a smaller $\RossbyL = \RoSmall$ [Fig.~\ref{fig:pairs_R2_components}(b)].
This is a well-known feature of rotating turbulence, previously evidenced in the Eulerian framework~\cite{Mininni2012, Delache2014, Vallefuoco2018}.

\smallskip

\paragraph{Effect of the pair initial orientation.}%
\label{par:effect_of_orientation}

In an anisotropic flow, the initial orientation $\rvecUnit$ of the Lagrangian pairs is also expected to have an impact on their separation rate, at least at short times following their release.
As introduced in Sec.~\ref{sub:lagrangian_tracking}, we consider here particle pairs initially oriented with the Cartesian axes, i.e.\ $\delta\xp_0 = r_0 \sqrt{2} \, \rvecUnit$ with $\rvecUnit \in \{\evec_x, \evec_y, \evec_z\}$.

\begin{figure}[tb]
  \centering
  \includegraphics[width = \textwidth]{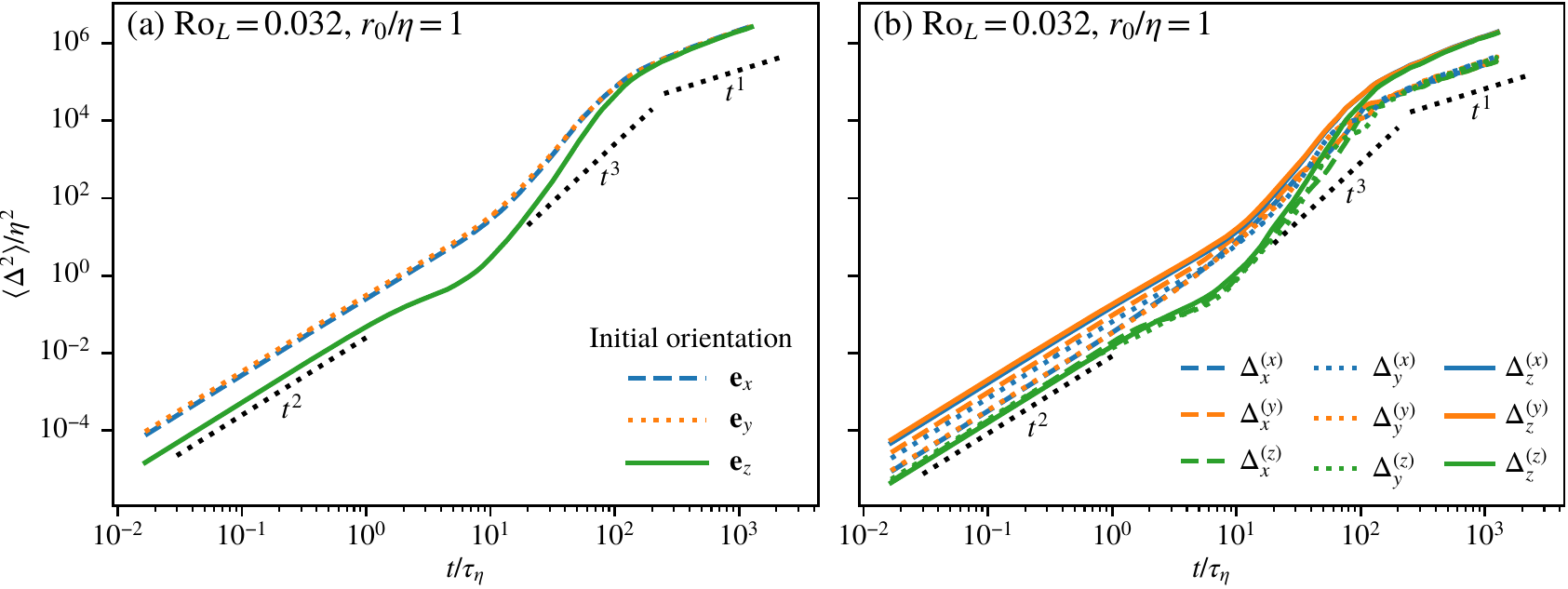}
  \caption{%
    Effect of the initial particle pair orientation on relative dispersion.
    (a) Time evolution of the total mean-squared separation $\Lmean{\DeltaFull^2}$ for pairs oriented along the $x$, $y$ and $z$ directions (dashed, dotted and solid lines, respectively).
    (b) Time evolution of the mean-squared separation $\Lmean{[\DeltaIJ{i}{j}]^2}$ along direction $i$ for pairs initially oriented along direction $j$.
    In all cases, the initial scale is $r_0 = \eta$ and the Rossby number is $\RossbyL = \RoSmall$ (Run 4).
  }%
  \label{fig:pairs_R2_oriented}
\end{figure}

We first consider the total separation $\DeltaFull$ instead of the component-wise separation $\Delta_i$.
As seen in Fig.~\ref{fig:pairs_R2_oriented}(a), in the ballistic regime and at high rotation rate,
pairs initially aligned with the rotation axis take considerably longer to separate than those initially oriented horizontally.
The effect of anisotropy is visible over a long time span, and disappears only when the diffusive regime is reached, i.e.\ when the memory of the initial condition has been lost.
As before, the observed short-time behavior can be linked to the Eulerian velocity structure functions.
Indeed, in the ballistic regime, $2 \Lmean{\DeltaFull^2}(t) \approx S_2^{(i)}(r_0) \, t^2$, where $S_2^{(i)}(r) = \Emean{[\uvec(\xvec + r \evec_i) - \uvec(\xvec)]^2}$ is the second-order structure function along direction $i$.
The slower separation rate for particles aligned with the rotation axis can therefore be attributed to the quasi-two-dimensionalization of the flow by rotation, which implies that the flow is nearly invariant along the rotation axis.
Another consequence of this short-time effect is that, at intermediate times, the separation rate is strongly accelerated for particles initially aligned with the rotation axis [Fig.~\ref{fig:pairs_R2_oriented}(a)].
This is due to the fact that these slow pairs need to catch up with faster ones such that, at long times, they all reach the diffusive regime.
This also explains the accelerated separation rate observed in Fig.~\ref{fig:pairs_R2}(a) at intermediate times for the highest rotation rate.

Figure~\ref{fig:pairs_R2_oriented}(b) shows the time evolution of the mean-squared component-wise separation of pairs conditioned on their initial orientation, $\Lmean{[\DeltaIJ{i}{j}]^2}$, where $\DeltaIJ{i}{j}(t) = |\delta X_i(t) - \delta X_i(0)| / \sqrt{2}$ such that $\delta \xp(0)$ is aligned with the Cartesian direction $\evec_j$.
In the ballistic regime, the following ordering holds:
$
\Lmean{[\DeltaIJ{x}{z}]^2} \sim
\Lmean{[\DeltaIJ{y}{z}]^2} \sim
\Lmean{[\DeltaIJ{z}{z}]^2} <
\Lmean{[\DeltaIJ{x}{x}]^2} \sim
\Lmean{[\DeltaIJ{y}{y}]^2} <
\Lmean{[\DeltaIJ{x}{y}]^2} \sim
\Lmean{[\DeltaIJ{y}{x}]^2} <
\Lmean{[\DeltaIJ{z}{x}]^2} \sim
\Lmean{[\DeltaIJ{z}{y}]^2}$.
Recalling that in this regime $2 \Lmean{[\DeltaIJ{i}{j}]^2}(t) \approx \StwoIJ{i}{j}(r) \, t^2$, where $\StwoIJ{i}{j}(r) = \Emean{[u_i(\xvec + r \evec_j) - u_i(\xvec)]^2}$ -- which reduces to $\Emean{(\partial_ju_i)^2}r^2$ in the limit of small $r$ -- this ordering is equivalent to the following one:
\begin{equation}
\Emean{(\partial_zu_x)^2} \sim \Emean{(\partial_zu_z)^2} < \Emean{(\partial_xu_x)^2} < \Emean{(\partial_xu_y)^2} < \Emean{(\partial_xu_z)^2} \label{eq:struct_fun}
\end{equation}
given the fact that, by symmetry in the horizontal plane, $\Emean{(\partial_zu_x)^2}=\Emean{(\partial_zu_y)^2}$, $\Emean{(\partial_xu_z)^2}=\Emean{(\partial_yu_z)^2}$, $\Emean{(\partial_yu_x)^2}=\Emean{(\partial_xu_y)^2}$, and $\Emean{(\partial_xu_x)^2}=\Emean{(\partial_yu_y)^2}$.
Equation (\ref{eq:struct_fun}) shows that the spatial variations in the vertical direction are the weakest, as expected in rotating turbulence.
The horizontal longitudinal second-order structure functions are larger than their transverse counterparts, as is the case in homogeneous and isotropic turbulence. The vertical velocity component displays the largest spatial variations, in the horizontal direction.
This ordering allows to interpret the short-time behavior of the mean-squared separations $\Delta_i$ for pairs initially randomly oriented [Fig.~\ref{fig:pairs_R2_components}(b)].
In Batchelor's regime, Fig.~\ref{fig:pairs_R2_oriented}(b) shows that pairs oriented horizontally at the initial time diffuse faster in the vertical direction than in the horizontal one [$\Lmean{(\DeltaIJ{z}{x,y})^2} > \Lmean{(\DeltaIJ{x,y}{x,y})^2}$], while those with an initial vertical orientation diffuse at comparable speeds in all directions [$\Lmean{(\Delta_{x,y}^{(z)})^2} \sim \Lmean{(\Delta_z^{(z)})^2}$].
Randomly-oriented particle pairs are therefore expected to diffuse faster in the $z$ direction, which is precisely the behavior observed in Fig.~\ref{fig:pairs_R2_components}(b).

\subsection{Lagrangian tetrads}%
\label{sub:dispersion_tetrads}

We investigate now the dispersion of Lagrangian tetrads in homogeneous rotating turbulence.
Tetrads encode additional information compared to that provided by particle pairs.
In particular, they allow to quantify the deformation of fluid elements.
In anisotropic flows, a preferential alignment of this deformation along different spatial directions is also expected.

\subsubsection{Tetrad volume}

\begin{figure}[tb]
  \centering
  \includegraphics[width = \textwidth]{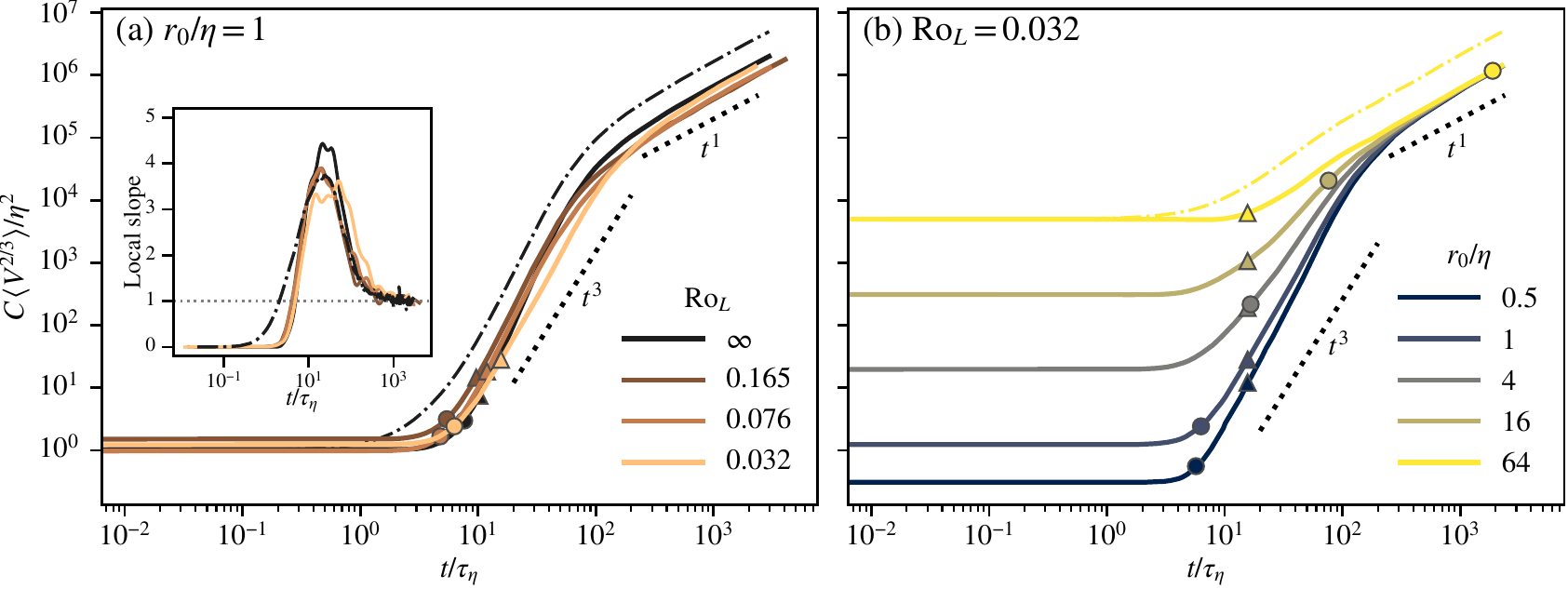}
  \caption{%
    Time evolution of the mean-squared tetrad size $\Lmean{V^{2/3}}$ estimated from their volume $V$ (solid lines).
    Curves are premultiplied by $C = 3^{2/3} \approx 2.08$.
    For comparison purposes, the normalized mean-squared tetrad gyration radius $\Lmean{R^2} / 3\eta^2$ is plotted for selected cases (dash-dotted lines).
    Markers are the same as in Fig.~\ref{fig:pairs_R2}.
    (a)~Tetrads of initial scale $r_0 = \eta$ and different rotation rates.
    Inset: local slope of the curves in the main panel.
    The horizontal dotted line indicates the scaling $t^1$.
    (b)~Tetrads of different initial scales $r_0$ under intense rotation ($\RossbyL = \RoSmall$).
  }%
  \label{fig:tetrads_volume}
\end{figure}

As opposed to particle pairs and triads, Lagrangian tetrads (and more generally clusters of $n\ge 4$ particles) have a volume $V$ that is generally non-zero.
The average volume-based tetrad size $\Lmean{V^{2/3}}$
is plotted in Fig.~\ref{fig:tetrads_volume} for different rotation rates and different initial tetrad sizes $r_0$
(the initial volume of the regular tetrahedrons herein considered is $V_0 = r_0^3 / 3$).
For reference, we also plot in dash-dotted lines the (compensated) mean-squared gyration radius $\Lmean{R^2} / 3$ of the tetrads, defined in Eq.~\eqref{eq:def_R2}, for selected cases.
Compensated this way, we have verified that the tetrad gyration radius superposes exactly to the pair gyration radius (Fig.~\ref{fig:pairs_R2}) at all times.
In fact, it can be shown that the compensated squared gyration radius $\Lmean{R^2} / (n - 1)$ is exactly the same for different cluster sizes $n$ (see Appendix~\ref{app:gyration_radius}).
This supports the idea that, while being very appropriate for estimating the shape and volume of Lagrangian objects, tetrads do not provide additional information on the linear growth of fluid blobs compared to particle pairs.

Interestingly, as seen in Fig.~\ref{fig:tetrads_volume}, the tetrad volume not only stays nearly constant at short times, but it does so for a considerably longer time than the gyration radius.
This is confirmed by the inset of Fig.~\ref{fig:tetrads_volume}(a), where the local slopes of the different curves are represented.
This slower change of volume
may be explained by the effect of incompressibility, which imposes a hard constraint $\dd V / \dd t = 0$ for sufficiently small tetrads (small enough so that velocity field is smooth at their scale).
Apart from this difference, the average evolution of the tetrad volume displays features qualitatively similar to the mean gyration radius, with an accelerated growth rate at intermediate times and an asymptotic long-time regime compatible with Taylor's diffusive regime, $\langle V^{2/3} \rangle \sim t$ [see in particular the inset of Fig.~\ref{fig:tetrads_volume}(a)].
This degree of similarity is also observed in the effect of the rotation rate, which
adds a time delay, in units of $\tau_\eta$, before the tetrad volume starts increasing.

\begin{figure}[tb]
  \centering
  \includegraphics[width = \textwidth]{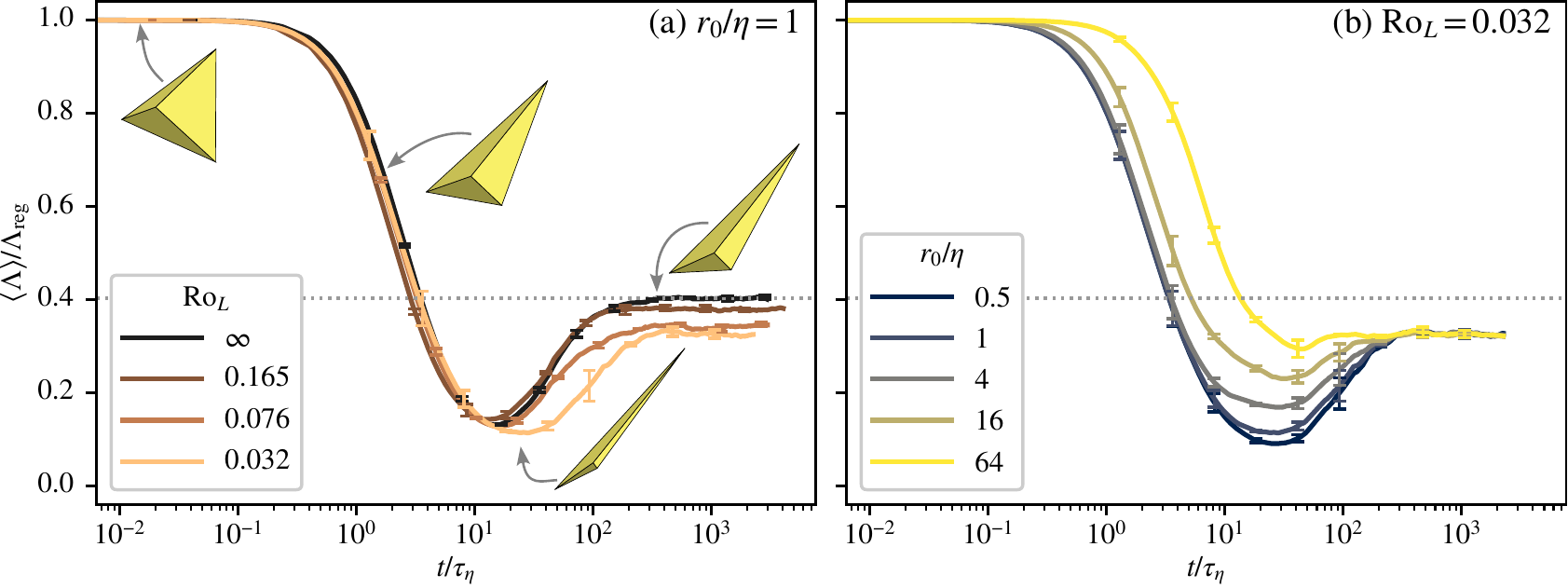}
  \caption{%
    Time evolution of the mean tetrad shape parameter $\Lmean{\Lambda} = \Lmean{V^{2/3} / R^2}$.
    Curves are normalized by the upper bound of the shape parameter, $\LambdaRegular = 3^{-5/3} \approx 6.24$, associated with regular tetrads.
    (a)~Tetrads of initial scale $r_0 = \eta$ and different rotation rates.
    (b)~Tetrads of different initial scales $r_0$ under intense rotation ($\RossbyL = \RoSmall$, Run 4).
    Error bars are obtained as described in the caption of Fig.~\ref{fig:pairs_R2}.
    The horizontal dotted line represents the asymptotic diffusive regime in HIT~\cite{Hackl2011}.
  }%
  \label{fig:tetrads_Lambda_mean}
\end{figure}

\subsubsection{Tetrad shape}

We now focus on the time evolution of the tetrad shape. A convenient way to characterize it is to measure
the non-dimensional parameter $\Lambda = V^{2/3} / R^2$ \cite{Hackl2011}, which is zero for sheet-like or needle-like structures
and reaches a maximal value of $\LambdaRegular = 3^{-5/3} \approx 0.16$ for perfectly regular tetrads (see Sec.~\ref{sub:geometric_definitions}).
The average value of $\Lambda$ is plotted in Fig.~\ref{fig:tetrads_Lambda_mean} for tetrads of different initial scales $r_0$ and for different rotation rates.
In all cases, the mean shape parameter $\Lmean{\Lambda}$ first drops from the initial value $\LambdaRegular = 3^{-5/3}$ associated with a regular tetrad.
The precise nature of this decay is likely a response to our choice of the initial tetrad shape.
The minimal value of $\Lmean{\Lambda}$, corresponding to a maximal average distortion of the structures, occurs at $t \approx (10-30)\tau_\eta$.
At this point, for $r_0 = \eta$, the average size of the tetrads is about $3\eta$ (see Fig.~\ref{fig:tetrads_volume}), at the low end of the inertial range.
The same behavior was observed in HIT by \citet{Pumir2000}, by using other indicators.
As seen in Fig.~\ref{fig:tetrads_Lambda_mean}(b), in rotating turbulence the minimal value of $\Lmean{\Lambda}$ significantly decreases as $r_0$ decreases.
The same behavior has been observed in isotropic turbulence, and attributed to the increasing influence of dissipative scale motions~\cite{Hackl2011}.

During this initial stage, rotation delays the time at which distortion is maximal (consistently with the delay observed for dispersion) and increases the strength of this deformation [Fig.~\ref{fig:tetrads_Lambda_mean}(a)].
In the presence of rotation, tetrads are therefore more likely to have sides of disparate lengths at $t \approx (10-30)\tau_\eta$, i.e.\ at the end of Batchelor's regime.
In other words, at this time rotation allows some pairs to be much further apart than others.
This is consistent with the results of Sec.~\ref{sub:dispersion_pairs} [Fig.~\ref{fig:pairs_R2_oriented}(a)].

At later times, the mean shape parameter $\Lmean{\Lambda}$ relaxes in the isotropic case towards the diffusive value $\Lambda^\infty \approx 0.0645$ ($\Lambda^\infty/\LambdaRegular\approx 0.40$, dotted horizontal lines), first estimated by \citet{Hackl2011}.
Figure~\ref{fig:tetrads_Lambda_mean}(a) shows that this asymptotic limit departs from the isotropic value under rotation, and decreases at decreasing Rossby number.
This means that, in the diffusive regime, tetrads become more and more elongated as the rotation rate increases, as qualitatively seen in Fig.~\ref{fig:visualisation}(b).
Their asymptotic shape is naturally independent of the initial conditions [Fig.~\ref{fig:tetrads_Lambda_mean}(b)].


\begin{figure}[tb]
  \centering
  \includegraphics[width = \textwidth]{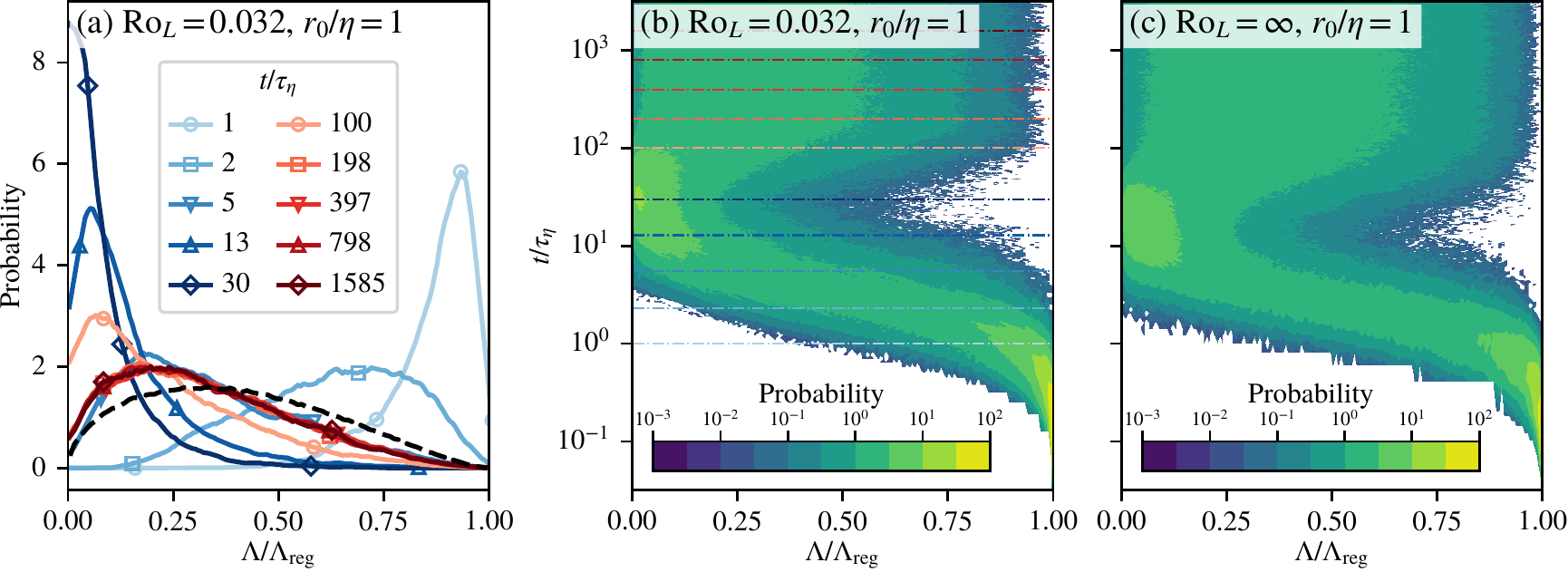}
  \caption{%
    Probability density function (PDF) of the tetrad shape parameter $\Lambda$.
    (a) PDF of $\Lambda$ at different times under intense rotation ($\RossbyL = \RoSmall$, Run 4), for tetrads of initial scale $r_0 = \eta$.
        Dashed black line: PDF of $\Lambda$ in the isotropic diffusive regime obtained from Monte Carlo simulations.
    (b) Alternative representation of the PDF of $\Lambda$ at different times, for the same case shown in (a).
        Each horizontal cut of the contour plot represents the PDF at a given time.
        Dash-dotted horizontal lines correspond to the different times plotted in (a).
    (c) Similar to (b) but for the non-rotating run (Run 1).
  }%
  \label{fig:tetrads_Lambda_pdf}
\end{figure}

A more complete description of the tetrad deformation is provided by the shape distribution.
The probability density functions (PDFs) of $\Lambda$ are plotted in Fig.~\ref{fig:tetrads_Lambda_pdf} at different times, for the strongly-rotating (a-b) and the non-rotating (c) cases.
Note that the two-dimensional representation in panels (b-c) is such that each horizontal cut corresponds to an instantaneous PDF at a given time.
At $t = 0$, since the tetrads are initially regular, the PDFs are described by a delta function centered at the maximal value $\Lambda = \LambdaRegular$ (not shown here).
The distributions, initially peaked near this value, quickly shift towards smaller values of $\Lambda$ at times $t \approx (10-30)\tau_\eta$, in agreement with Fig.~\ref{fig:tetrads_Lambda_mean}.
Interestingly, the distribution is strongly peaked when distortion is maximal, meaning that a very large fraction of the tetrads are then subject to this strong deformation.
The distribution then broadens and relaxes at long times towards an asymptotic diffusive state.
These observations apply both to rotating and to isotropic turbulence, and are consistent with previous results in the latter case~\cite{Hackl2011}.
In particular, in the absence of rotation, we have checked (not shown) that the PDF relaxes towards the diffusive state expected for isotropic turbulence.
This state, represented by a dashed line in Fig.~\ref{fig:tetrads_Lambda_pdf}(a), is not recovered in the presence of rotation, highlighting the persistent effect of rotation on the deformation of Lagrangian blobs.

\begin{figure}[tb]
  \centering
  \includegraphics[width = \textwidth]{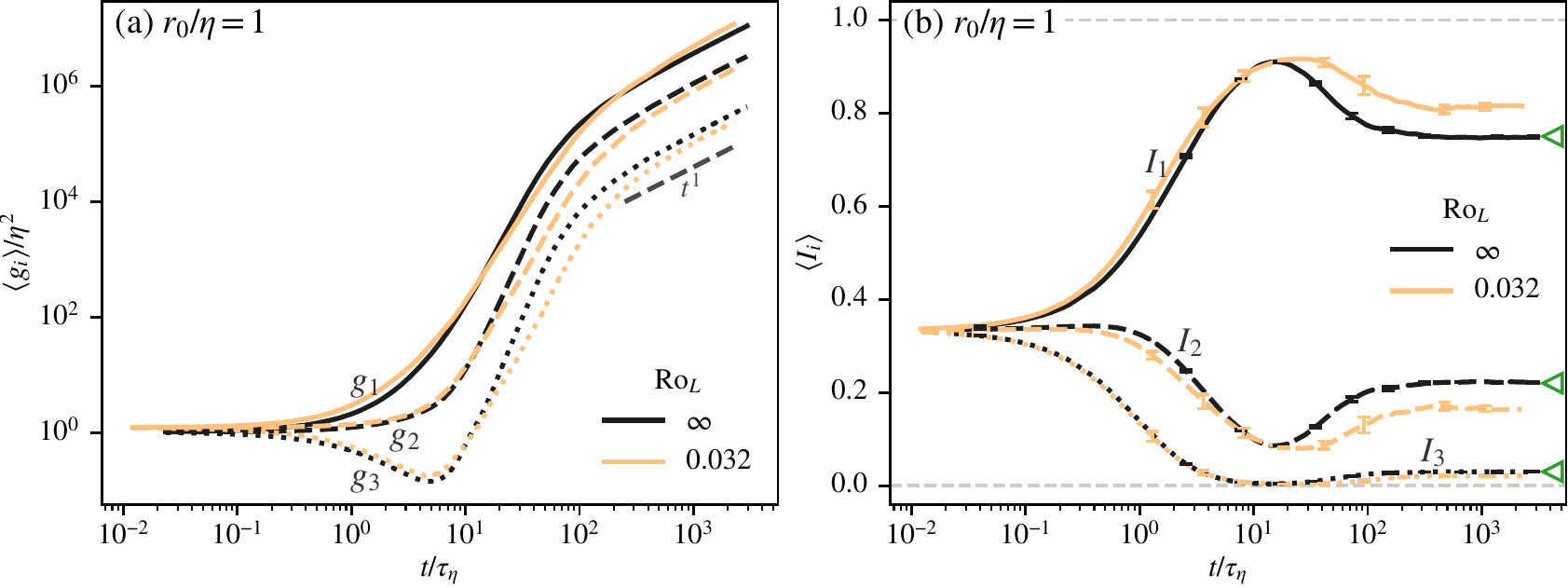}
  \caption{%
    Average statistics of tetrad shape eigenvalues.
    (a) Average tetrad eigenvalues $\Lmean{g_i}$.
    (b) Average tetrad shape parameters $\Lmean{I_i}$, with $I_i = g_i / R^2$.
    Solid lines, $i = 1$; dashed lines, $i = 2$; dotted lines, $i = 3$.
    In each panel, the non-rotating and strongly-rotating cases are shown.
    In all cases, the initial tetrad scale is $r_0 = \eta$.
    In (b), triangles mark the expected diffusive values in HIT.
    Error bars are obtained as described in the caption of Fig.~\ref{fig:pairs_R2}.
  }%
  \label{fig:tetrads_eigenvalues_mean}
\end{figure}

A tetrad is a three-dimensional structure and its shape cannot therefore be fully represented by a single parameter.
A more complete characterization of the shape is provided by the eigenvalues $g_i$ of the moment-of-inertia tensor, defined in Sec.~\ref{sub:geometric_definitions}.
Recall that $g_1\ge g_2\ge g_3$ by convention, that the sum of these quantities is the squared tetrad size $R^2$, and that their product is, up to a scaling factor, the squared tetrad volume $V^2$.
Figure~\ref{fig:tetrads_eigenvalues_mean}(a) shows the time dependence of the mean values of $g_i$.
The three eigenvalues remain constant at short times, before a shape relaxation from the chosen initial condition occurs.
In the long-time limit, the three eigenvalues exhibit a clear diffusive regime, $\Lmean{g_i} \sim t$, like their sum $\Lmean{R^2}$ (Fig.~\ref{fig:pairs_R2}).

We now focus on the non-dimensional shape factors $I_i = \gval{i} / R^2$, commonly used to characterize the tetrad deformation in HIT~\cite{Pumir2000, Biferale2005b, Luthi2007, Xu2008, Hackl2011}.
We recall that only two of these three parameters are independent since their sum is, by definition, equal to unity.
The evolution of their respective mean values is plotted in Fig.~\ref{fig:tetrads_eigenvalues_mean}(b) in non-rotating and in strongly-rotating turbulence.
At very short times, all eigenvalues are equal for regular tetrads, namely $g_i = r_0^2$ and $I_i = 1/3$.
At later times, their mean values significantly depart from this initial value, consistently with what was observed for the volume-based shape parameter $\Lambda$ (Fig.~\ref{fig:tetrads_Lambda_mean}).
At $t \sim \tau_\eta$ the stretching eigenvalue $g_1$ is much larger than the compressive one $g_3$, suggesting that tetrads are already strongly-flattened at times comparable to the dissipative time scale.
Consistently with Fig.~\ref{fig:tetrads_Lambda_mean}, the mean tetrad distortion is maximal when its size is of order $3\eta$ ($t \approx (10-30)\tau_\eta$), and rotation delays the time at which this occurs. The mean value of $I_3$ is then very close to zero, which indicates that the structures are extremely flattened (sheet-like).

The three eigenvalues relax at long time towards asymptotic values characterizing the tetrad mean shape in the diffusive regime.
In the isotropic case, these values are in agreement with those calculated for a Gaussian distribution [see green triangles in Fig.~\ref{fig:tetrads_eigenvalues_mean}(b)]: $\Lmean{I_1}^\infty \approx 0.75$, $\Lmean{I_2}^\infty \approx 0.22$ and $\Lmean{I_1}^\infty \approx 0.03$.
This corresponds to very elongated structures~\cite{Pumir2000, Biferale2005b, Luthi2007, Xu2008, Hackl2011} ($\Lmean{I_1}^\infty\approx 25 \Lmean{I_3}^\infty$).
In the presence of rotation, the asymptotic shapes are even more flattened: for $\RossbyL = \RoSmall$, we get $\Lmean{I_1}^\infty \approx 0.81$, $\Lmean{I_2}^\infty \approx 0.17$ and $\Lmean{I_1}^\infty \approx 0.02$, therefore $\Lmean{I_1}^\infty \approx 40\Lmean{I_3}^\infty$.
The ratio between the two largest shape factors is also larger in the presence of rotation: $\Lmean{I_1}^\infty/\Lmean{I_2}^\infty$ is $\approx 3.4$ in THI, and $\approx 4.8$ for $\RossbyL = \RoSmall$.
The large faces of the tetrads are therefore more distorted in the strongly-rotating case.

The anisotropy of rotating turbulence, and in particular the formation of large-scale columnar vortices aligned with the rotation axis, is also expected to have a signature on the direction in which tetrads are stretched or compressed.
We therefore focus now on the orientation of these structures with respect to the rotation axis.

\subsubsection{Tetrad orientation}

\begin{figure}[tb]
  \centering
  \includegraphics[width = \textwidth]{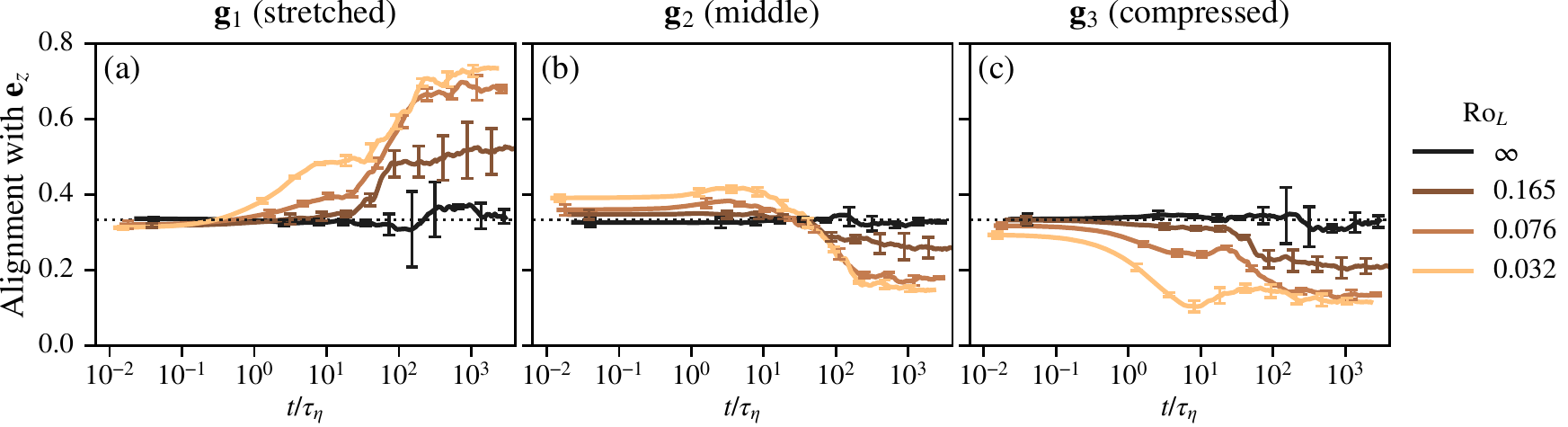}
  \caption{%
    Mean alignment
    $\Alignment_i= \langle (\gvec{i}\cdot\evec_z)^2 \rangle$
    between the moment-of-inertia tensor eigenvectors
    and the Cartesian axis $\evec_z$, aligned with the rotation axis.
    Horizontal dotted lines represent the isotropic value $\Alignment_i = 1/3$.
    In all cases, the initial tetrad scale is $r_0 = \eta$.
    Error bars are obtained as described in the caption of Fig.~\ref{fig:pairs_R2}.
  }%
  \label{fig:tetrads_orientations}
\end{figure}

The tetrad orientation with respect to the rotation axis can be characterized by measuring the variance of the scalar product between the unit vectors $\gvec{i}$ and $\evec_z$, $\Alignment_i\equiv \langle (\gvec{i}\cdot\evec_z)^2 \rangle$.
Both vectors are preferentially collinear if $\Alignment_i>1/3$, normal to each other if $\Alignment_i<1/3$, and oriented independently if $\Alignment_i=1/3$.
The time evolution of the mean alignment $\Alignment_i$ is shown in Fig.~\ref{fig:tetrads_orientations} for different rotation rates.
In isotropic turbulence, $\gvec{i}$ is expected to be randomly oriented and one expects $\Alignment_i$ to be equal to $1/3$ for all $i\in \{1,2,3\}$ and at any time.
This is confirmed at sufficiently long times (black curves), while $\Alignment_i$ weakly departs from this value at shorter times.
This may be attributed to an imperfect statistical convergence due to the finite number of tetrads tracked in the simulations, or to the short-lived presence of anisotropic structures in the statistically-isotropic simulations.
In any case, this departure is negligible in comparison with the preferential alignment visible at long times in the rotating flows.

In the presence of rotation, $\Alignment_i$ departs from the value $1/3$ at times shorter than $10\tau_\eta$.
The three indicators evolve in time until they reach their asymptotic values $\Alignment_i^\infty$.
For the lowest Rossby number considered, $\Alignment_1^\infty\approx 0.70$, $\Alignment_2^\infty\approx 0.17$, and $\Alignment_3^\infty\approx 0.13$.
This means that the largest dimension of the tetrads is preferentially vertical (aligned with the rotation axis), and the two other ones are preferentially horizontal (perpendicular to this axis).
This is illustrated in Fig.~\ref{fig:visualisation}(b), in which most of the tetrads are very elongated and their maximal dimensions are oriented vertically.
  Before reaching the discussed asymptotic states, the tetrad alignments follow a temporal dynamics which is not necessarily monotonic. This is in particular the case for the alignment of $\gvec{2}$ with the rotation axis [Fig.~\ref{fig:tetrads_orientations}(b)]:
  at short times, $\gvec{2}$ is slightly preferentially collinear with the rotation vector $\RotationVec$, whereas in the asymptotic state, both vectors are preferentially perpendicular to each other.
  Such a behavior can be interpreted by recalling that, at short times, the deformation of isotropic tetrads is strongly influenced by the local rate of strain \cite{Pumir2013}. As shown in~\cite{Naso2019}, in rotating turbulence the eigenvector associated with the largest (positive) eigenvalue of the strain is weakly perpendicular to $\RotationVec$, the one associated to the smallest (negative) eigenvalue is normal to it, and the eigenvector associated to the intermediate (positive) eigenvalue is strongly collinear to $\RotationVec$. This explains why, at short times, $\gvec{1}$ is preferentially slighly perpendicular to $\RotationVec$, whereas $\gvec{2}$ and $\gvec{3}$ respectively tend to be collinear and normal to $\RotationVec$. At long times, the behaviors reflected in Fig.~\ref{fig:tetrads_orientations} can be interpreted by using the results obtained in Sec.~\ref{sub:dispersion_pairs} for pairs: as shown in Fig.~\ref{fig:pairs_R2_components}, at long times the vertical relative dispersion is faster than the horizontal one, which explains why tetrads tend to align vertically in the flow.

\subsection{Lagrangian triads}%
\label{sub:dispersion_triads}

We finally investigate the dispersion and distortion of triads, vertices of initially-equilateral triangles.
The full description of the geometric features of a three-dimensional flow requires to consider clusters of at least four particles, but investigating the shape of triads allows to extract general considerations on the effect of the number of particles in a cluster on its shape factors~\cite{Hackl2011}.

\begin{figure}[tb]
  \centering
  \includegraphics[width = \textwidth]{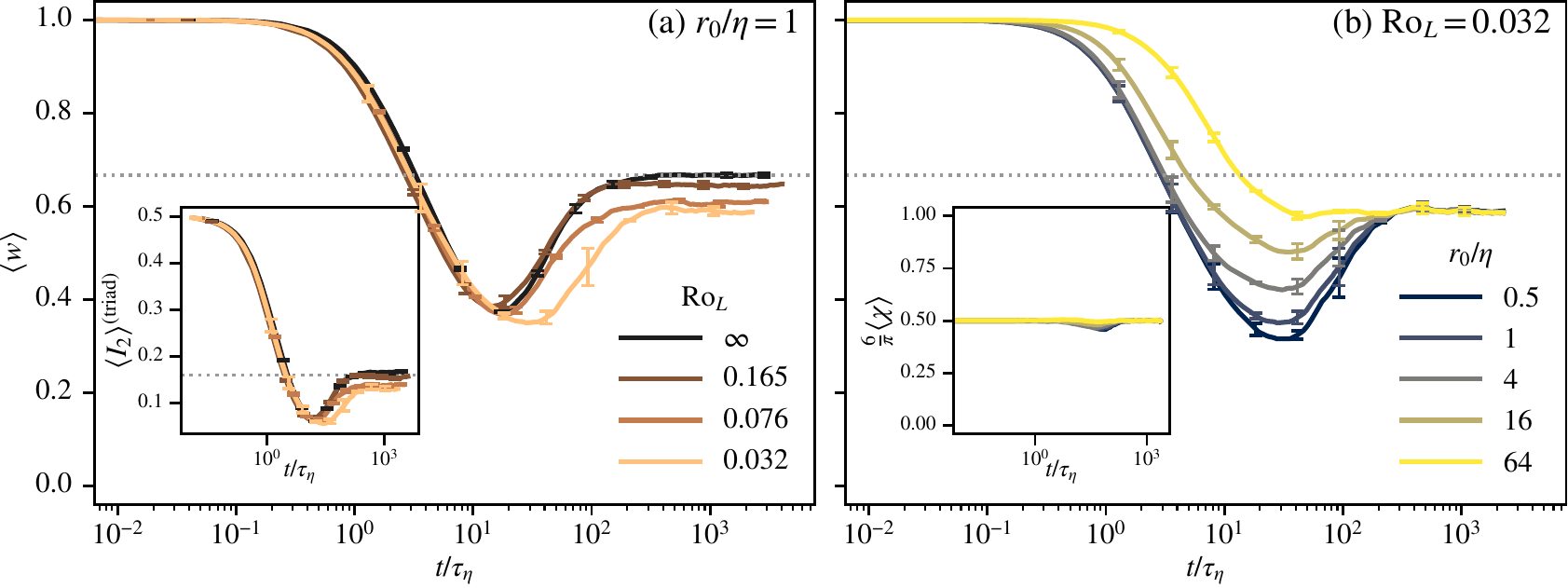}
  \caption{%
    Main panels: time evolution of the mean triad shape parameter $\Lmean{w}$.
    (a)~Triads of initial scale $r_0 = \eta$ and different rotation rates.
    (b)~Triads of different initial scales $r_0$ under intense rotation ($\RossbyL = \RoSmall$, Run 4).
    Insets: (a) mean shape factor $\Lmean{\ITriad{2}}$;
    (b) Euler angle $\Lmean{\chi}$ normalized by its upper bound $\chi_{\text{max}} = \pi / 6$.
    Error bars are obtained as described in the caption of Fig.~\ref{fig:pairs_R2}.
    Horizontal dotted gray lines represent the asymptotic diffusive values in high Reynolds number HIT~\cite{Pumir2000,Hackl2011}, $\Lmean{w} = 2/3$, $\Lmean{\ITriad{2}} \approx 0.16$.
  }%
  \label{fig:triads_w_mean}
\end{figure}

As mentioned earlier for tetrads and shown in Appendix~\ref{app:gyration_radius}, the statistics of the gyration radius $R$ of the considered regular triads is equivalent to that of pairs (Fig.~\ref{fig:pairs_R2}), so we focus here on their distortion.
A common measure of deformation of triads is the non-dimensional shape parameter $w = A / \AreaRegular = 4A / (\sqrt{3} R^2)$~\cite{Pumir1998, Shraiman1998, Castiglione2001, Hackl2011}, where $A$ is the triangle area and $\AreaRegular = \sqrt{3} R^2 / 4$ is the area of an equilateral triangle with the same gyration radius.
This parameter takes the values $w = 0$ and $w = 1$ in the respective limiting cases of a collinear triangle with zero area and of an equilateral triangle.
As seen in the main panels of Fig.~\ref{fig:triads_w_mean}, the average value of $w$ evolves similarly to that of $\Lambda$ for tetrads (Fig.~\ref{fig:tetrads_Lambda_mean}), indicating that, just like the latter, triads become strongly deformed at intermediate times, before relaxing towards an asymptotic shape factor at long times.
A more straightforward comparison with tetrads can be performed by considering the eigenvalues of the triangle's moment-of-inertia tensor.
The inset of Fig.~\ref{fig:triads_w_mean}(a) shows the time dependence of the average of the smallest normalized eigenvalue $\ITriad{2} = \gvalTriad{2} / \left(\gvalTriad{1} + \gvalTriad{2}\right)$ (recall that, by definition, $\ITriad{1} = 1 - \ITriad{2}$).
The general trends of $\Lmean{\ITriad{2}}(t)$ are the same as those of $\Lmean{w}(t)$.

In the isotropic flow, the asymptotic values of the triad shape factors display a very satisfactory agreement with their expected values~\cite{Pumir2000,Hackl2011}, $\Lmean{w}^\infty\approx 2/3$ and $\Lmean{\ITriad{2}}^\infty\approx 0.16$, as shown by the horizontal dotted lines in Fig.~\ref{fig:triads_w_mean}(a).
These asymptotic values decrease at increasing rotation rate, meaning that, in the long-time diffusive regime, triads are more distorted in rotating turbulence.
They seem to reach a limit value for the two smallest $\RossbyL$, for which $\Lmean{w}^\infty\approx 0.6$ and $\Lmean{\ITriad{2}}^\infty\approx 0.13$.


Figure~\ref{fig:triads_w_mean}(b) shows that the dependence of the triads shape factor on the initial size $r_0$ is the same as for tetrads [Fig.~\ref{fig:tetrads_Lambda_mean}(b)]), namely the maximal mean distortion increases at decreasing $r_0$.
This trend is observed for all considered rotation rates.
Finally, the inset of the same figure shows the time dependence of the Euler angle $\chi$~\cite{Shraiman1998, Castiglione2001}, which provides information on triangle's symmetry.
In the present simulations $\Lmean{\chi}$ is equal to $\pi/12$ at all times, for all initial scales $r_0$ and all the considered rotation rates.
Note that at $t=0$, the value of $\chi$ for equilateral triangles is not well-posed.
As particles move away from this initial condition, the fact that $\Lmean{\chi}  \approx\pi/12$ means that the triangles do not seem to become more or less isosceles on average.
The value $\pi/12$ is also the expected diffusive limit in HIT~\cite{Pumir1998,Hackl2011}.
It is reached here for all rotation rates.

\begin{figure}[tb]
  \centering
  \includegraphics[width = \textwidth]{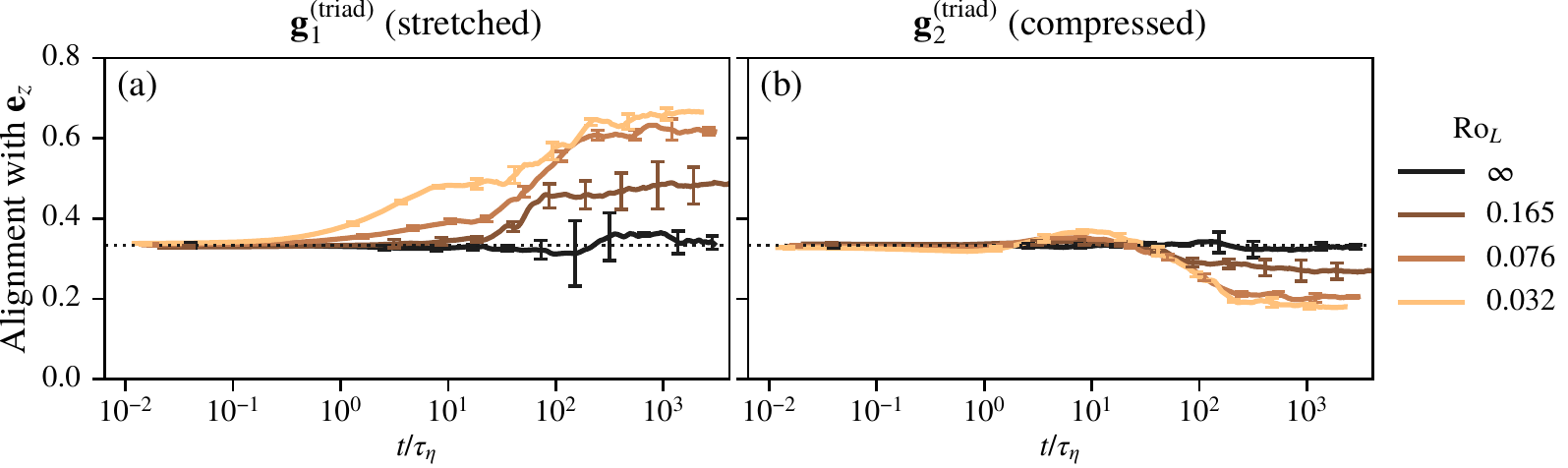}
  \caption{%
    Mean alignment $\Lmean{(\gvecTriad{i} \cdot \evec_z)^2}$ between triangle eigendirections $\gvecTriad{i}$ and the Cartesian axis $\evec_z$.
    Horizontal dotted lines represent the isotropic value $1/3$.
    In all cases, the initial triangle scale is $r_0 = \eta$.
    Error bars are obtained as described in the caption of Fig.~\ref{fig:pairs_R2}.
  }%
  \label{fig:triads_orientations}
\end{figure}

We finally consider the preferential orientation of the triangle eigenvectors $\gvecTriad{i}$ relative to the global rotation axis.
As mentioned in Sec.~\ref{sub:geometric_definitions}, $g_3 = 0$ for triangles, and therefore only $\gvecTriad{1}$ and $\gvecTriad{2}$ are relevant to describe their deformation axes.
These two eigenvectors are respectively associated with the directions in which a triangle is most elongated or compressed.
The average orientation of these eigendirections with the rotation axis is plotted in Fig.~\ref{fig:triads_orientations}.
Overall, the results are quantitatively similar to those obtained for tetrads (Fig.~\ref{fig:tetrads_orientations}).
As with tetrads, these can be summarized by a preferential stretching of triangles along the axis of rotation.
Besides, it is interesting to note that the compressive direction $\gvecTriad{2}$ displays a behavior that is most similar to the intermediate eigendirection $\gvec{2}$ for tetrads, instead of being similar to the compressive tetrad direction $\gvec{3}$.
For the smallest Rossby number and in the long-time limit, $\langle (\gvecTriad{2}\cdot\evec_z)^2 \rangle \approx 0.19$ for triads, whereas for tetrads $\langle (\gvec{2}\cdot\evec_z)^2 \rangle \approx 0.16$ and $\langle (\gvec{3}\cdot\evec_z)^2 \rangle \approx 0.12$.
These results may be intuitively understood by considering the four triangular faces of an extremely flattened tetrad.
In this case, the tetrad can be seen as composed of two nearly-parallel triangles of large area, connected by other two triangles of negligible area.
The large triangles are expected to have the same eigendirections as the tetrad, so that they are approximately orthogonal to the tetrad compressive direction $\gvec{3}$.
The added contributions of the large triangles may explain the similarity between the orientation of triad and tetrad eigenvectors.

\section{Conclusions and discussion} \label{sec:concl}

We have investigated the Lagrangian properties of homogeneous rotating turbulence using direct numerical simulations.
We have first characterized single-particle Lagrangian velocity autocorrelations, finding a strong modification of this quantity compared to the isotropic (rotationless) case.
In particular, the autocorrelation of the horizontal velocity components (perpendicular to the rotation axis) displays oscillations
with a period close to the rotation one, $\Trot = 2\pi / \OmegaRot$.
This is likely explained by the effect of large-scale cyclonic and anticyclonic vortex columns that rotate at a rate comparable to $\OmegaRot$.
Meanwhile, the vertical velocity component (parallel to the rotation axis) remains correlated for
a longer time.
This time is an increasing function of the rotation rate.

We have then focused on the statistics of the size and shape of two-, three- and four-particle Lagrangian clusters.
For the initially isotropic triads and tetrads considered in the present work,
the size statistics are, at first order, completely described by the mean-squared separation of particle pairs.
By subtracting the initial separation, we have recovered in all cases a clear ballistic regime at short times, expected to occur as long as particles have memory of their relative initial velocity.
At later times, an accelerated separation regime is observed.
While qualitatively compatible, this regime cannot be attributed to Richardson's explosive regime, as the Lagrangian scale separation of our simulations is not sufficient.
Finally, at long times, the diffusive regime first predicted by Taylor is observed.
In this regime, and in the presence of rotation, the diffusion is faster in the vertical direction than in the horizontal one, as a result of the slower decorrelation of the vertical velocity.

The initial orientation of a particle pair has an important influence on their dispersion at short and intermediate times.
In particular, pairs oriented horizontally at the release time diffuse faster in the vertical direction than in the horizontal one, while pairs initially vertical diffuse at speeds comparable in both directions.
As a consequence, at short times randomly-oriented pairs diffuse faster in the vertical direction than in the horizontal one.
These results are the signature of the anisotropy of Eulerian second-order structure functions.

We have then focused on the deformation and preferential orientation of Lagrangian tetrads and triads.
In both cases, the clusters undergo a strong flattening at intermediate times, before their shape statistics relax towards a diffusive regime.
This is consistent with previous observations in isotropic turbulence.
Rotation delays the time at which this distortion is maximal, while also increasing its strength.
It also increases the shape distortion in the asymptotic regime, in which tetrads transported by rotating turbulence are more flattened and have more elongated large faces than in HIT.
Furthermore, and consistently with our observations for pairs, clusters tend to preferentially stretch along the rotation axis.
The effect of rotation on triads is similar: in the presence of rotation these clusters are more distorted, both in the intermediate and in the asymptotic regimes, and they are preferentially aligned vertically.




We have dealt with Lagrangian dispersion in the ideal framework of homogeneous rotating turbulence.
Other effects which are relevant for geophysical applications, in particular vertical density stratification and confinement, are absent in this setting.
They are both expected to suppress the transport of Lagrangian tracers along the vertical direction~\cite{Buaria2020a,Liechtenstein2006,DelCastello2011a}, therefore one may expect some of the conclusions of the present work to change under those conditions.
It would be interesting in the future to perform numerical studies of multi-particle dispersion in confined rotating geometries, to better characterize the effect of confinement and of the formation of Ekman layers on Lagrangian transport.
An example of such idealized setting is the rotating channel flow geometry, previously used to investigate Eulerian features of confined rotating flows~\cite{Godeferd1999}.
Multi-particle statistics could also be investigated in homogeneous turbulence subject to both rotation and stratification.

\begin{acknowledgments}
  This work was supported by the IDEXLYON project (contract ANR-16-IDEX-0005) under University of Lyon auspices.
  Simulations were carried out using the facilities of the PMCS2I (École Centrale de Lyon).
\end{acknowledgments}

\appendix

\section{Equivalence between gyration radii of $n$-particle clusters}\label{app:gyration_radius}

In the following we consider the compensated gyration radius $\Rcomp_n \equiv R_n / \sqrt{n - 1}$ of $n$-particle clusters.
The purpose of this section is to show that, given a set of $n$ particles, the compensated squared gyration radius $\Rcomp_n^2$ of the resulting cluster is equal to $\Rcomp_2^2$ averaged among all particle pairs that compose the cluster.
That is,
\begin{equation}
  \Rcomp^2_n = \Lmean{\Rcomp_2^2}
  \quad \longleftrightarrow \quad
  R^2_n = (n - 1) \Lmean{R^2_2}
  \quad \text{for } n \ge 3,
\end{equation}
where averages are performed over all possible particle pair combinations within the set of $n$ particles.

Using the first definition in Eq.~\eqref{eq:def_R2}, we write the compensated gyration radius of an $n$-particle cluster as
\begin{equation}
  \Rcomp^2_n
  = \frac{1}{2n (n - 1)} \sum_{i = 1}^n \sum_{j = 1}^n \DistSq_{ij}
  = \frac{1}{n (n - 1)} \sum_{i = 1}^{n - 1} \sum_{j = i + 1}^n \DistSq_{ij},
\end{equation}
where $\DistSq_{ij} = |\xp_i - \xp_j|$ is the distance between two particles.
In the last equality we have used the $i \leftrightarrow j$ symmetry of $\DistSq_{ij}$, as well as the fact that $\DistSq_{ii} = 0$ for all $i$.
Noting that the last double-sum is over the $n (n - 1) / 2$ possible combinations of $\Dist_{ij}$ within the $n$-particle cluster, we can alternatively write
\begin{equation}
  \label{eq:app:gyration_radius}
  \Rcomp^2_n
  = \frac{1}{n (n - 1)} \frac{n (n - 1)}{2} \Lmean{\DistSq_{ij}}
  = \frac{1}{2} \Lmean{\DistSq_{ij}}
  = \Lmean{R^2_2},
\end{equation}
where $\Lmean{\DistSq_{ij}}$ is the average squared distance among all particle pairs composing the $n$-particle cluster.
In the last equality, we have used the fact that the squared gyration radius of a single particle pair $(\xp_i, \xp_j)$ is precisely $R^2_2 = \DistSq_{ij} / 2$.

Equation~\eqref{eq:app:gyration_radius} thus shows that the compensated squared gyration radius $R^2_n / (n - 1)$ of an $n$-particle cluster is equal to the mean squared gyration radius $\Lmean{R^2_2}$ of all pairs composing the $n$-particle cluster.
This supports the intuition that tracking pairs of particles in turbulent flows is sufficient to obtain a full description of the linear growth of a patch of fluid.
Specifically in the context of the present work, this means that gyration radius statistics are expected to be equivalent for tetrads, triads and pairs -- which we have verified in our data.

\bibliography{multiparticles}

\begin{thebibliography}{57}%
\makeatletter
\providecommand \@ifxundefined [1]{%
 \@ifx{#1\undefined}
}%
\providecommand \@ifnum [1]{%
 \ifnum #1\expandafter \@firstoftwo
 \else \expandafter \@secondoftwo
 \fi
}%
\providecommand \@ifx [1]{%
 \ifx #1\expandafter \@firstoftwo
 \else \expandafter \@secondoftwo
 \fi
}%
\providecommand \natexlab [1]{#1}%
\providecommand \enquote  [1]{``#1''}%
\providecommand \bibnamefont  [1]{#1}%
\providecommand \bibfnamefont [1]{#1}%
\providecommand \citenamefont [1]{#1}%
\providecommand \href@noop [0]{\@secondoftwo}%
\providecommand \href [0]{\begingroup \@sanitize@url \@href}%
\providecommand \@href[1]{\@@startlink{#1}\@@href}%
\providecommand \@@href[1]{\endgroup#1\@@endlink}%
\providecommand \@sanitize@url [0]{\catcode `\\12\catcode `\$12\catcode
  `\&12\catcode `\#12\catcode `\^12\catcode `\_12\catcode `\%12\relax}%
\providecommand \@@startlink[1]{}%
\providecommand \@@endlink[0]{}%
\providecommand \url  [0]{\begingroup\@sanitize@url \@url }%
\providecommand \@url [1]{\endgroup\@href {#1}{\urlprefix }}%
\providecommand \urlprefix  [0]{URL }%
\providecommand \Eprint [0]{\href }%
\providecommand \doibase [0]{https://doi.org/}%
\providecommand \selectlanguage [0]{\@gobble}%
\providecommand \bibinfo  [0]{\@secondoftwo}%
\providecommand \bibfield  [0]{\@secondoftwo}%
\providecommand \translation [1]{[#1]}%
\providecommand \BibitemOpen [0]{}%
\providecommand \bibitemStop [0]{}%
\providecommand \bibitemNoStop [0]{.\EOS\space}%
\providecommand \EOS [0]{\spacefactor3000\relax}%
\providecommand \BibitemShut  [1]{\csname bibitem#1\endcsname}%
\let\auto@bib@innerbib\@empty
\bibitem [{\citenamefont {Yeung}(2002)}]{Yeung2002}%
  \BibitemOpen
  \bibfield  {author} {\bibinfo {author} {\bibfnamefont {P.~K.}\ \bibnamefont
  {Yeung}},\ }\bibfield  {title} {\bibinfo {title} {Lagrangian investigations
  of turbulence},\ }\href
  {https://doi.org/10.1146/annurev.fluid.34.082101.170725} {\bibfield
  {journal} {\bibinfo  {journal} {Annu. Rev. Fluid Mech.}\ }\textbf {\bibinfo
  {volume} {34}},\ \bibinfo {pages} {115} (\bibinfo {year} {2002})}\BibitemShut
  {NoStop}%
\bibitem [{\citenamefont {Toschi}\ and\ \citenamefont
  {Bodenschatz}(2009)}]{Toschi2009}%
  \BibitemOpen
  \bibfield  {author} {\bibinfo {author} {\bibfnamefont {F.}~\bibnamefont
  {Toschi}}\ and\ \bibinfo {author} {\bibfnamefont {E.}~\bibnamefont
  {Bodenschatz}},\ }\bibfield  {title} {\bibinfo {title} {Lagrangian properties
  of particles in turbulence},\ }\href
  {https://doi.org/10.1146/annurev.fluid.010908.165210} {\bibfield  {journal}
  {\bibinfo  {journal} {Annu. Rev. Fluid Mech.}\ }\textbf {\bibinfo {volume}
  {41}},\ \bibinfo {pages} {375} (\bibinfo {year} {2009})}\BibitemShut
  {NoStop}%
\bibitem [{\citenamefont {La~Porta}\ \emph {et~al.}(2001)\citenamefont
  {La~Porta}, \citenamefont {Voth}, \citenamefont {Crawford}, \citenamefont
  {Alexander},\ and\ \citenamefont {Bodenschatz}}]{LaPorta2001}%
  \BibitemOpen
  \bibfield  {author} {\bibinfo {author} {\bibfnamefont {A.}~\bibnamefont
  {La~Porta}}, \bibinfo {author} {\bibfnamefont {G.~A.}\ \bibnamefont {Voth}},
  \bibinfo {author} {\bibfnamefont {A.~M.}\ \bibnamefont {Crawford}}, \bibinfo
  {author} {\bibfnamefont {J.}~\bibnamefont {Alexander}},\ and\ \bibinfo
  {author} {\bibfnamefont {E.}~\bibnamefont {Bodenschatz}},\ }\bibfield
  {title} {\bibinfo {title} {Fluid particle accelerations in fully developed
  turbulence},\ }\href {https://doi.org/10.1038/35059027} {\bibfield  {journal}
  {\bibinfo  {journal} {Nature}\ }\textbf {\bibinfo {volume} {409}},\ \bibinfo
  {pages} {1017} (\bibinfo {year} {2001})}\BibitemShut {NoStop}%
\bibitem [{\citenamefont {Mordant}\ \emph {et~al.}(2002)\citenamefont
  {Mordant}, \citenamefont {Delour}, \citenamefont {L{\'e}veque}, \citenamefont
  {Arn{\'e}odo},\ and\ \citenamefont {Pinton}}]{Mordant2002}%
  \BibitemOpen
  \bibfield  {author} {\bibinfo {author} {\bibfnamefont {N.}~\bibnamefont
  {Mordant}}, \bibinfo {author} {\bibfnamefont {J.}~\bibnamefont {Delour}},
  \bibinfo {author} {\bibfnamefont {E.}~\bibnamefont {L{\'e}veque}}, \bibinfo
  {author} {\bibfnamefont {A.}~\bibnamefont {Arn{\'e}odo}},\ and\ \bibinfo
  {author} {\bibfnamefont {J.-F.}\ \bibnamefont {Pinton}},\ }\bibfield  {title}
  {\bibinfo {title} {Long {{Time Correlations}} in {{Lagrangian Dynamics}}: {{A
  Key}} to {{Intermittency}} in {{Turbulence}}},\ }\href
  {https://doi.org/10.1103/PhysRevLett.89.254502} {\bibfield  {journal}
  {\bibinfo  {journal} {Phys. Rev. Lett.}\ }\textbf {\bibinfo {volume} {89}},\
  \bibinfo {pages} {254502} (\bibinfo {year} {2002})}\BibitemShut {NoStop}%
\bibitem [{\citenamefont {Arn{\`e}odo}\ \emph {et~al.}(2008)\citenamefont
  {Arn{\`e}odo}, \citenamefont {Benzi}, \citenamefont {Berg}, \citenamefont
  {Biferale}, \citenamefont {Bodenschatz}, \citenamefont {Busse}, \citenamefont
  {Calzavarini}, \citenamefont {Castaing}, \citenamefont {Cencini},
  \citenamefont {Chevillard}, \citenamefont {Fisher}, \citenamefont {Grauer},
  \citenamefont {Homann}, \citenamefont {Lamb}, \citenamefont {Lanotte},
  \citenamefont {L{\'e}v{\`e}que}, \citenamefont {L{\"u}thi}, \citenamefont
  {Mann}, \citenamefont {Mordant}, \citenamefont {M{\"u}ller}, \citenamefont
  {Ott}, \citenamefont {Ouellette}, \citenamefont {Pinton}, \citenamefont
  {Pope}, \citenamefont {Roux}, \citenamefont {Toschi}, \citenamefont {Xu},\
  and\ \citenamefont {Yeung}}]{Arneodo2008}%
  \BibitemOpen
  \bibfield  {author} {\bibinfo {author} {\bibfnamefont {A.}~\bibnamefont
  {Arn{\`e}odo}}, \bibinfo {author} {\bibfnamefont {R.}~\bibnamefont {Benzi}},
  \bibinfo {author} {\bibfnamefont {J.}~\bibnamefont {Berg}}, \bibinfo {author}
  {\bibfnamefont {L.}~\bibnamefont {Biferale}}, \bibinfo {author}
  {\bibfnamefont {E.}~\bibnamefont {Bodenschatz}}, \bibinfo {author}
  {\bibfnamefont {A.}~\bibnamefont {Busse}}, \bibinfo {author} {\bibfnamefont
  {E.}~\bibnamefont {Calzavarini}}, \bibinfo {author} {\bibfnamefont
  {B.}~\bibnamefont {Castaing}}, \bibinfo {author} {\bibfnamefont
  {M.}~\bibnamefont {Cencini}}, \bibinfo {author} {\bibfnamefont
  {L.}~\bibnamefont {Chevillard}}, \bibinfo {author} {\bibfnamefont {R.~T.}\
  \bibnamefont {Fisher}}, \bibinfo {author} {\bibfnamefont {R.}~\bibnamefont
  {Grauer}}, \bibinfo {author} {\bibfnamefont {H.}~\bibnamefont {Homann}},
  \bibinfo {author} {\bibfnamefont {D.}~\bibnamefont {Lamb}}, \bibinfo {author}
  {\bibfnamefont {A.~S.}\ \bibnamefont {Lanotte}}, \bibinfo {author}
  {\bibfnamefont {E.}~\bibnamefont {L{\'e}v{\`e}que}}, \bibinfo {author}
  {\bibfnamefont {B.}~\bibnamefont {L{\"u}thi}}, \bibinfo {author}
  {\bibfnamefont {J.}~\bibnamefont {Mann}}, \bibinfo {author} {\bibfnamefont
  {N.}~\bibnamefont {Mordant}}, \bibinfo {author} {\bibfnamefont {W.-C.}\
  \bibnamefont {M{\"u}ller}}, \bibinfo {author} {\bibfnamefont
  {S.}~\bibnamefont {Ott}}, \bibinfo {author} {\bibfnamefont {N.~T.}\
  \bibnamefont {Ouellette}}, \bibinfo {author} {\bibfnamefont {J.-F.}\
  \bibnamefont {Pinton}}, \bibinfo {author} {\bibfnamefont {S.~B.}\
  \bibnamefont {Pope}}, \bibinfo {author} {\bibfnamefont {S.~G.}\ \bibnamefont
  {Roux}}, \bibinfo {author} {\bibfnamefont {F.}~\bibnamefont {Toschi}},
  \bibinfo {author} {\bibfnamefont {H.}~\bibnamefont {Xu}},\ and\ \bibinfo
  {author} {\bibfnamefont {P.~K.}\ \bibnamefont {Yeung}} (\bibinfo
  {collaboration} {International Collaboration for Turbulence Research}),\
  }\bibfield  {title} {\bibinfo {title} {Universal {{Intermittent Properties}}
  of {{Particle Trajectories}} in {{Highly Turbulent Flows}}},\ }\href
  {https://doi.org/10.1103/PhysRevLett.100.254504} {\bibfield  {journal}
  {\bibinfo  {journal} {Phys. Rev. Lett.}\ }\textbf {\bibinfo {volume} {100}},\
  \bibinfo {pages} {254504} (\bibinfo {year} {2008})}\BibitemShut {NoStop}%
\bibitem [{\citenamefont {Sawford}(2001)}]{Sawford2001}%
  \BibitemOpen
  \bibfield  {author} {\bibinfo {author} {\bibfnamefont {B.}~\bibnamefont
  {Sawford}},\ }\bibfield  {title} {\bibinfo {title} {Turbulent relative
  dispersion},\ }\href@noop {} {\bibfield  {journal} {\bibinfo  {journal}
  {Annu. Rev. Fluid Mech.}\ }\textbf {\bibinfo {volume} {33}},\ \bibinfo
  {pages} {289} (\bibinfo {year} {2001})}\BibitemShut {NoStop}%
\bibitem [{\citenamefont {Richardson}(1926)}]{Richardson1926}%
  \BibitemOpen
  \bibfield  {author} {\bibinfo {author} {\bibfnamefont {L.~F.}\ \bibnamefont
  {Richardson}},\ }\bibfield  {title} {\bibinfo {title} {Atmospheric diffusion
  shown on a distance-neighbour graph},\ }\href@noop {} {\bibfield  {journal}
  {\bibinfo  {journal} {Proc. R. Soc. Lond. Ser. A}\ }\textbf {\bibinfo
  {volume} {110}},\ \bibinfo {pages} {709} (\bibinfo {year}
  {1926})}\BibitemShut {NoStop}%
\bibitem [{\citenamefont {Salazar}\ and\ \citenamefont
  {Collins}(2009)}]{Salazar2009}%
  \BibitemOpen
  \bibfield  {author} {\bibinfo {author} {\bibfnamefont {J.~P. L.~C.}\
  \bibnamefont {Salazar}}\ and\ \bibinfo {author} {\bibfnamefont {L.~R.}\
  \bibnamefont {Collins}},\ }\bibfield  {title} {\bibinfo {title} {Two-particle
  dispersion in isotropic turbulent flows},\ }\href
  {https://doi.org/10.1146/annurev.fluid.40.111406.102224} {\bibfield
  {journal} {\bibinfo  {journal} {Annu. Rev. Fluid Mech.}\ }\textbf {\bibinfo
  {volume} {41}},\ \bibinfo {pages} {405} (\bibinfo {year} {2009})}\BibitemShut
  {NoStop}%
\bibitem [{\citenamefont {Batchelor}(1950)}]{Batchelor1950}%
  \BibitemOpen
  \bibfield  {author} {\bibinfo {author} {\bibfnamefont {G.~K.}\ \bibnamefont
  {Batchelor}},\ }\bibfield  {title} {\bibinfo {title} {The application of the
  similarity theory of turbulence to atmospheric diffusion},\ }\href
  {https://doi.org/10.1002/qj.49707632804} {\bibfield  {journal} {\bibinfo
  {journal} {Q.J.R. Meteorol. Soc.}\ }\textbf {\bibinfo {volume} {76}},\
  \bibinfo {pages} {133} (\bibinfo {year} {1950})}\BibitemShut {NoStop}%
\bibitem [{\citenamefont {Taylor}(1922)}]{Taylor1922}%
  \BibitemOpen
  \bibfield  {author} {\bibinfo {author} {\bibfnamefont {G.~I.}\ \bibnamefont
  {Taylor}},\ }\bibfield  {title} {\bibinfo {title} {Diffusion by continuous
  movements},\ }\href {https://doi.org/10.1112/plms/s2-20.1.196} {\bibfield
  {journal} {\bibinfo  {journal} {Proc. Lond. Math. Soc.}\ }\textbf {\bibinfo
  {volume} {s2-20}},\ \bibinfo {pages} {196} (\bibinfo {year}
  {1922})}\BibitemShut {NoStop}%
\bibitem [{\citenamefont {Chertkov}\ \emph {et~al.}(1999)\citenamefont
  {Chertkov}, \citenamefont {Pumir},\ and\ \citenamefont
  {Shraiman}}]{Chertkov1999}%
  \BibitemOpen
  \bibfield  {author} {\bibinfo {author} {\bibfnamefont {M.}~\bibnamefont
  {Chertkov}}, \bibinfo {author} {\bibfnamefont {A.}~\bibnamefont {Pumir}},\
  and\ \bibinfo {author} {\bibfnamefont {B.~I.}\ \bibnamefont {Shraiman}},\
  }\bibfield  {title} {\bibinfo {title} {Lagrangian tetrad dynamics and the
  phenomenology of turbulence},\ }\href {https://doi.org/10.1063/1.870101}
  {\bibfield  {journal} {\bibinfo  {journal} {Phys. Fluids}\ }\textbf {\bibinfo
  {volume} {11}},\ \bibinfo {pages} {2394} (\bibinfo {year}
  {1999})}\BibitemShut {NoStop}%
\bibitem [{\citenamefont {Pumir}\ \emph {et~al.}(2000)\citenamefont {Pumir},
  \citenamefont {Shraiman},\ and\ \citenamefont {Chertkov}}]{Pumir2000}%
  \BibitemOpen
  \bibfield  {author} {\bibinfo {author} {\bibfnamefont {A.}~\bibnamefont
  {Pumir}}, \bibinfo {author} {\bibfnamefont {B.~I.}\ \bibnamefont
  {Shraiman}},\ and\ \bibinfo {author} {\bibfnamefont {M.}~\bibnamefont
  {Chertkov}},\ }\bibfield  {title} {\bibinfo {title} {Geometry of {{Lagrangian
  Dispersion}} in {{Turbulence}}},\ }\href
  {https://doi.org/10.1103/PhysRevLett.85.5324} {\bibfield  {journal} {\bibinfo
   {journal} {Phys. Rev. Lett.}\ }\textbf {\bibinfo {volume} {85}},\ \bibinfo
  {pages} {5324} (\bibinfo {year} {2000})}\BibitemShut {NoStop}%
\bibitem [{\citenamefont {Biferale}\ \emph {et~al.}(2005)\citenamefont
  {Biferale}, \citenamefont {Boffetta}, \citenamefont {Celani}, \citenamefont
  {Devenish}, \citenamefont {Lanotte},\ and\ \citenamefont
  {Toschi}}]{Biferale2005b}%
  \BibitemOpen
  \bibfield  {author} {\bibinfo {author} {\bibfnamefont {L.}~\bibnamefont
  {Biferale}}, \bibinfo {author} {\bibfnamefont {G.}~\bibnamefont {Boffetta}},
  \bibinfo {author} {\bibfnamefont {A.}~\bibnamefont {Celani}}, \bibinfo
  {author} {\bibfnamefont {B.~J.}\ \bibnamefont {Devenish}}, \bibinfo {author}
  {\bibfnamefont {A.}~\bibnamefont {Lanotte}},\ and\ \bibinfo {author}
  {\bibfnamefont {F.}~\bibnamefont {Toschi}},\ }\bibfield  {title} {\bibinfo
  {title} {Multiparticle dispersion in fully developed turbulence},\ }\href
  {https://doi.org/10.1063/1.2130751} {\bibfield  {journal} {\bibinfo
  {journal} {Phys. Fluids}\ }\textbf {\bibinfo {volume} {17}},\ \bibinfo
  {pages} {111701} (\bibinfo {year} {2005})}\BibitemShut {NoStop}%
\bibitem [{\citenamefont {Hackl}\ \emph {et~al.}(2011)\citenamefont {Hackl},
  \citenamefont {Yeung},\ and\ \citenamefont {Sawford}}]{Hackl2011}%
  \BibitemOpen
  \bibfield  {author} {\bibinfo {author} {\bibfnamefont {J.~F.}\ \bibnamefont
  {Hackl}}, \bibinfo {author} {\bibfnamefont {P.~K.}\ \bibnamefont {Yeung}},\
  and\ \bibinfo {author} {\bibfnamefont {B.~L.}\ \bibnamefont {Sawford}},\
  }\bibfield  {title} {\bibinfo {title} {Multi-particle and tetrad statistics
  in numerical simulations of turbulent relative dispersion},\ }\href
  {https://doi.org/10.1063/1.3586803} {\bibfield  {journal} {\bibinfo
  {journal} {Phys. Fluids}\ }\textbf {\bibinfo {volume} {23}},\ \bibinfo
  {pages} {065103} (\bibinfo {year} {2011})}\BibitemShut {NoStop}%
\bibitem [{\citenamefont {Pumir}\ \emph {et~al.}(2013)\citenamefont {Pumir},
  \citenamefont {Bodenschatz},\ and\ \citenamefont {Xu}}]{Pumir2013}%
  \BibitemOpen
  \bibfield  {author} {\bibinfo {author} {\bibfnamefont {A.}~\bibnamefont
  {Pumir}}, \bibinfo {author} {\bibfnamefont {E.}~\bibnamefont {Bodenschatz}},\
  and\ \bibinfo {author} {\bibfnamefont {H.}~\bibnamefont {Xu}},\ }\bibfield
  {title} {\bibinfo {title} {Tetrahedron deformation and alignment of perceived
  vorticity and strain in a turbulent flow},\ }\href
  {https://doi.org/10.1063/1.4795547} {\bibfield  {journal} {\bibinfo
  {journal} {Phys. Fluids}\ }\textbf {\bibinfo {volume} {25}},\ \bibinfo
  {pages} {035101} (\bibinfo {year} {2013})}\BibitemShut {NoStop}%
\bibitem [{\citenamefont {Castiglione}\ and\ \citenamefont
  {Pumir}(2001)}]{Castiglione2001}%
  \BibitemOpen
  \bibfield  {author} {\bibinfo {author} {\bibfnamefont {P.}~\bibnamefont
  {Castiglione}}\ and\ \bibinfo {author} {\bibfnamefont {A.}~\bibnamefont
  {Pumir}},\ }\bibfield  {title} {\bibinfo {title} {Evolution of triangles in a
  two-dimensional turbulent flow},\ }\href
  {https://doi.org/10.1103/PhysRevE.64.056303} {\bibfield  {journal} {\bibinfo
  {journal} {Phys. Rev. E}\ }\textbf {\bibinfo {volume} {64}},\ \bibinfo
  {pages} {056303} (\bibinfo {year} {2001})}\BibitemShut {NoStop}%
\bibitem [{\citenamefont {Hopfinger}\ \emph {et~al.}(1982)\citenamefont
  {Hopfinger}, \citenamefont {Browand},\ and\ \citenamefont
  {Gagne}}]{Hopfinger1982}%
  \BibitemOpen
  \bibfield  {author} {\bibinfo {author} {\bibfnamefont {E.~J.}\ \bibnamefont
  {Hopfinger}}, \bibinfo {author} {\bibfnamefont {F.~K.}\ \bibnamefont
  {Browand}},\ and\ \bibinfo {author} {\bibfnamefont {Y.}~\bibnamefont
  {Gagne}},\ }\bibfield  {title} {\bibinfo {title} {Turbulence and waves in a
  rotating tank},\ }\href {https://doi.org/10.1017/S0022112082003462}
  {\bibfield  {journal} {\bibinfo  {journal} {J. Fluid Mech.}\ }\textbf
  {\bibinfo {volume} {125}},\ \bibinfo {pages} {505} (\bibinfo {year}
  {1982})}\BibitemShut {NoStop}%
\bibitem [{\citenamefont {Smith}\ and\ \citenamefont
  {Waleffe}(1999)}]{Smith1999}%
  \BibitemOpen
  \bibfield  {author} {\bibinfo {author} {\bibfnamefont {L.~M.}\ \bibnamefont
  {Smith}}\ and\ \bibinfo {author} {\bibfnamefont {F.}~\bibnamefont
  {Waleffe}},\ }\bibfield  {title} {\bibinfo {title} {Transfer of energy to
  two-dimensional large scales in forced, rotating three-dimensional
  turbulence},\ }\href {https://doi.org/10.1063/1.870022} {\bibfield  {journal}
  {\bibinfo  {journal} {Phys. Fluids}\ }\textbf {\bibinfo {volume} {11}},\
  \bibinfo {pages} {1608} (\bibinfo {year} {1999})}\BibitemShut {NoStop}%
\bibitem [{\citenamefont {Moisy}\ \emph {et~al.}(2011)\citenamefont {Moisy},
  \citenamefont {Morize}, \citenamefont {Rabaud},\ and\ \citenamefont
  {Sommeria}}]{Moisy2011}%
  \BibitemOpen
  \bibfield  {author} {\bibinfo {author} {\bibfnamefont {F.}~\bibnamefont
  {Moisy}}, \bibinfo {author} {\bibfnamefont {C.}~\bibnamefont {Morize}},
  \bibinfo {author} {\bibfnamefont {M.}~\bibnamefont {Rabaud}},\ and\ \bibinfo
  {author} {\bibfnamefont {J.}~\bibnamefont {Sommeria}},\ }\bibfield  {title}
  {\bibinfo {title} {Decay laws, anisotropy and cyclone\textendash anticyclone
  asymmetry in decaying rotating turbulence},\ }\href
  {https://doi.org/10.1017/S0022112010003733} {\bibfield  {journal} {\bibinfo
  {journal} {J. Fluid Mech.}\ }\textbf {\bibinfo {volume} {666}},\ \bibinfo
  {pages} {5} (\bibinfo {year} {2011})}\BibitemShut {NoStop}%
\bibitem [{\citenamefont {Godeferd}\ and\ \citenamefont
  {Moisy}(2015)}]{Godeferd2015}%
  \BibitemOpen
  \bibfield  {author} {\bibinfo {author} {\bibfnamefont {F.~S.}\ \bibnamefont
  {Godeferd}}\ and\ \bibinfo {author} {\bibfnamefont {F.}~\bibnamefont
  {Moisy}},\ }\bibfield  {title} {\bibinfo {title} {Structure and dynamics of
  rotating turbulence: A review of recent experimental and numerical results},\
  }\href {https://doi.org/10.1115/1.4029006} {\bibfield  {journal} {\bibinfo
  {journal} {Appl. Mech. Rev.}\ }\textbf {\bibinfo {volume} {67}},\ \bibinfo
  {pages} {030802} (\bibinfo {year} {2015})}\BibitemShut {NoStop}%
\bibitem [{\citenamefont {Naso}(2015)}]{Naso2015}%
  \BibitemOpen
  \bibfield  {author} {\bibinfo {author} {\bibfnamefont {A.}~\bibnamefont
  {Naso}},\ }\bibfield  {title} {\bibinfo {title} {Cyclone-anticyclone
  asymmetry and alignment statistics in homogeneous rotating turbulence},\
  }\href {https://doi.org/10.1063/1.4914176} {\bibfield  {journal} {\bibinfo
  {journal} {Phys. Fluids}\ }\textbf {\bibinfo {volume} {27}},\ \bibinfo
  {pages} {035108} (\bibinfo {year} {2015})}\BibitemShut {NoStop}%
\bibitem [{\citenamefont {Liechtenstein}\ \emph {et~al.}(2006)\citenamefont
  {Liechtenstein}, \citenamefont {Godeferd},\ and\ \citenamefont
  {Cambon}}]{Liechtenstein2006}%
  \BibitemOpen
  \bibfield  {author} {\bibinfo {author} {\bibfnamefont {L.}~\bibnamefont
  {Liechtenstein}}, \bibinfo {author} {\bibfnamefont {F.}~\bibnamefont
  {Godeferd}},\ and\ \bibinfo {author} {\bibfnamefont {C.}~\bibnamefont
  {Cambon}},\ }\bibfield  {title} {\bibinfo {title} {The role of nonlinearity
  in turbulent diffusion models for stably stratified and rotating
  turbulence},\ }\href {https://doi.org/10.1016/j.ijheatfluidflow.2006.02.010}
  {\bibfield  {journal} {\bibinfo  {journal} {Int. J. Heat Fluid Flow}\
  }\textbf {\bibinfo {volume} {27}},\ \bibinfo {pages} {644} (\bibinfo {year}
  {2006})}\BibitemShut {NoStop}%
\bibitem [{\citenamefont {{van Aartrijk}}\ \emph {et~al.}(2008)\citenamefont
  {{van Aartrijk}}, \citenamefont {Clercx},\ and\ \citenamefont
  {Winters}}]{vanAartrijk2008}%
  \BibitemOpen
  \bibfield  {author} {\bibinfo {author} {\bibfnamefont {M.}~\bibnamefont {{van
  Aartrijk}}}, \bibinfo {author} {\bibfnamefont {H.~J.~H.}\ \bibnamefont
  {Clercx}},\ and\ \bibinfo {author} {\bibfnamefont {K.~B.}\ \bibnamefont
  {Winters}},\ }\bibfield  {title} {\bibinfo {title} {Single-particle,
  particle-pair, and multiparticle dispersion of fluid particles in forced
  stably stratified turbulence},\ }\href {https://doi.org/10.1063/1.2838593}
  {\bibfield  {journal} {\bibinfo  {journal} {Phys. Fluids}\ }\textbf {\bibinfo
  {volume} {20}},\ \bibinfo {pages} {025104} (\bibinfo {year}
  {2008})}\BibitemShut {NoStop}%
\bibitem [{\citenamefont {Cambon}\ \emph {et~al.}(2004)\citenamefont {Cambon},
  \citenamefont {Godeferd}, \citenamefont {Nicolleau},\ and\ \citenamefont
  {Vassilicos}}]{Cambon2004}%
  \BibitemOpen
  \bibfield  {author} {\bibinfo {author} {\bibfnamefont {C.}~\bibnamefont
  {Cambon}}, \bibinfo {author} {\bibfnamefont {F.~S.}\ \bibnamefont
  {Godeferd}}, \bibinfo {author} {\bibfnamefont {F.~C. G.~A.}\ \bibnamefont
  {Nicolleau}},\ and\ \bibinfo {author} {\bibfnamefont {J.~C.}\ \bibnamefont
  {Vassilicos}},\ }\bibfield  {title} {\bibinfo {title} {Turbulent diffusion in
  rapidly rotating flows with and without stable stratification},\ }\href
  {https://doi.org/10.1017/S0022112003007055} {\bibfield  {journal} {\bibinfo
  {journal} {J. Fluid Mech.}\ }\textbf {\bibinfo {volume} {499}},\ \bibinfo
  {pages} {231} (\bibinfo {year} {2004})}\BibitemShut {NoStop}%
\bibitem [{\citenamefont {Del~Castello}\ and\ \citenamefont
  {Clercx}(2011{\natexlab{a}})}]{DelCastello2011}%
  \BibitemOpen
  \bibfield  {author} {\bibinfo {author} {\bibfnamefont {L.}~\bibnamefont
  {Del~Castello}}\ and\ \bibinfo {author} {\bibfnamefont {H.~J.~H.}\
  \bibnamefont {Clercx}},\ }\bibfield  {title} {\bibinfo {title} {Lagrangian
  {{Acceleration}} of {{Passive Tracers}} in {{Statistically Steady Rotating
  Turbulence}}},\ }\href {https://doi.org/10.1103/PhysRevLett.107.214502}
  {\bibfield  {journal} {\bibinfo  {journal} {Phys. Rev. Lett.}\ }\textbf
  {\bibinfo {volume} {107}},\ \bibinfo {pages} {214502} (\bibinfo {year}
  {2011}{\natexlab{a}})}\BibitemShut {NoStop}%
\bibitem [{\citenamefont {Del~Castello}\ and\ \citenamefont
  {Clercx}(2011{\natexlab{b}})}]{DelCastello2011a}%
  \BibitemOpen
  \bibfield  {author} {\bibinfo {author} {\bibfnamefont {L.}~\bibnamefont
  {Del~Castello}}\ and\ \bibinfo {author} {\bibfnamefont {H.~J.~H.}\
  \bibnamefont {Clercx}},\ }\bibfield  {title} {\bibinfo {title} {Lagrangian
  velocity autocorrelations in statistically steady rotating turbulence},\
  }\href {https://doi.org/10.1103/PhysRevE.83.056316} {\bibfield  {journal}
  {\bibinfo  {journal} {Phys. Rev. E}\ }\textbf {\bibinfo {volume} {83}},\
  \bibinfo {pages} {056316} (\bibinfo {year} {2011}{\natexlab{b}})}\BibitemShut
  {NoStop}%
\bibitem [{\citenamefont {Alards}\ \emph {et~al.}(2017)\citenamefont {Alards},
  \citenamefont {Rajaei}, \citenamefont {Del~Castello}, \citenamefont {Kunnen},
  \citenamefont {Toschi},\ and\ \citenamefont {Clercx}}]{Alards2017}%
  \BibitemOpen
  \bibfield  {author} {\bibinfo {author} {\bibfnamefont {K.~M.~J.}\
  \bibnamefont {Alards}}, \bibinfo {author} {\bibfnamefont {H.}~\bibnamefont
  {Rajaei}}, \bibinfo {author} {\bibfnamefont {L.}~\bibnamefont
  {Del~Castello}}, \bibinfo {author} {\bibfnamefont {R.~P.~J.}\ \bibnamefont
  {Kunnen}}, \bibinfo {author} {\bibfnamefont {F.}~\bibnamefont {Toschi}},\
  and\ \bibinfo {author} {\bibfnamefont {H.~J.~H.}\ \bibnamefont {Clercx}},\
  }\bibfield  {title} {\bibinfo {title} {Geometry of tracer trajectories in
  rotating turbulent flows},\ }\href
  {https://doi.org/10.1103/PhysRevFluids.2.044601} {\bibfield  {journal}
  {\bibinfo  {journal} {Phys. Rev. Fluids}\ }\textbf {\bibinfo {volume} {2}},\
  \bibinfo {pages} {044601} (\bibinfo {year} {2017})}\BibitemShut {NoStop}%
\bibitem [{\citenamefont {Maity}\ \emph {et~al.}(2019)\citenamefont {Maity},
  \citenamefont {Govindarajan},\ and\ \citenamefont {Ray}}]{Maity2019}%
  \BibitemOpen
  \bibfield  {author} {\bibinfo {author} {\bibfnamefont {P.}~\bibnamefont
  {Maity}}, \bibinfo {author} {\bibfnamefont {R.}~\bibnamefont
  {Govindarajan}},\ and\ \bibinfo {author} {\bibfnamefont {S.~S.}\ \bibnamefont
  {Ray}},\ }\bibfield  {title} {\bibinfo {title} {Statistics of {{Lagrangian}}
  trajectories in a rotating turbulent flow},\ }\href
  {https://doi.org/10.1103/PhysRevE.100.043110} {\bibfield  {journal} {\bibinfo
   {journal} {Phys. Rev. E}\ }\textbf {\bibinfo {volume} {100}},\ \bibinfo
  {pages} {043110} (\bibinfo {year} {2019})}\BibitemShut {NoStop}%
\bibitem [{\citenamefont {Buaria}\ \emph {et~al.}(2020)\citenamefont {Buaria},
  \citenamefont {Pumir}, \citenamefont {Feraco}, \citenamefont {Marino},
  \citenamefont {Pouquet}, \citenamefont {Rosenberg},\ and\ \citenamefont
  {Primavera}}]{Buaria2020a}%
  \BibitemOpen
  \bibfield  {author} {\bibinfo {author} {\bibfnamefont {D.}~\bibnamefont
  {Buaria}}, \bibinfo {author} {\bibfnamefont {A.}~\bibnamefont {Pumir}},
  \bibinfo {author} {\bibfnamefont {F.}~\bibnamefont {Feraco}}, \bibinfo
  {author} {\bibfnamefont {R.}~\bibnamefont {Marino}}, \bibinfo {author}
  {\bibfnamefont {A.}~\bibnamefont {Pouquet}}, \bibinfo {author} {\bibfnamefont
  {D.}~\bibnamefont {Rosenberg}},\ and\ \bibinfo {author} {\bibfnamefont
  {L.}~\bibnamefont {Primavera}},\ }\bibfield  {title} {\bibinfo {title}
  {Single-particle {{Lagrangian}} statistics from direct numerical simulations
  of rotating-stratified turbulence},\ }\href
  {https://doi.org/10.1103/PhysRevFluids.5.064801} {\bibfield  {journal}
  {\bibinfo  {journal} {Phys. Rev. Fluids}\ }\textbf {\bibinfo {volume} {5}},\
  \bibinfo {pages} {064801} (\bibinfo {year} {2020})}\BibitemShut {NoStop}%
\bibitem [{\citenamefont {Pumir}(1998)}]{Pumir1998}%
  \BibitemOpen
  \bibfield  {author} {\bibinfo {author} {\bibfnamefont {A.}~\bibnamefont
  {Pumir}},\ }\bibfield  {title} {\bibinfo {title} {Structure of the
  three-point correlation function of a passive scalar in the presence of a
  mean gradient},\ }\href {https://doi.org/10.1103/PhysRevE.57.2914} {\bibfield
   {journal} {\bibinfo  {journal} {Phys. Rev. E}\ }\textbf {\bibinfo {volume}
  {57}},\ \bibinfo {pages} {2914} (\bibinfo {year} {1998})}\BibitemShut
  {NoStop}%
\bibitem [{\citenamefont {Shraiman}\ and\ \citenamefont
  {Siggia}(1998)}]{Shraiman1998}%
  \BibitemOpen
  \bibfield  {author} {\bibinfo {author} {\bibfnamefont {B.~I.}\ \bibnamefont
  {Shraiman}}\ and\ \bibinfo {author} {\bibfnamefont {E.~D.}\ \bibnamefont
  {Siggia}},\ }\bibfield  {title} {\bibinfo {title} {Anomalous scaling for a
  passive scalar near the {{Batchelor}} limit},\ }\href
  {https://doi.org/10.1103/PhysRevE.57.2965} {\bibfield  {journal} {\bibinfo
  {journal} {Phys. Rev. E}\ }\textbf {\bibinfo {volume} {57}},\ \bibinfo
  {pages} {2965} (\bibinfo {year} {1998})}\BibitemShut {NoStop}%
\bibitem [{\citenamefont {Bourgoin}\ \emph {et~al.}(2006)\citenamefont
  {Bourgoin}, \citenamefont {Ouellette}, \citenamefont {Xu}, \citenamefont
  {Berg},\ and\ \citenamefont {Bodenschatz}}]{Bourgoin2006}%
  \BibitemOpen
  \bibfield  {author} {\bibinfo {author} {\bibfnamefont {M.}~\bibnamefont
  {Bourgoin}}, \bibinfo {author} {\bibfnamefont {N.~T.}\ \bibnamefont
  {Ouellette}}, \bibinfo {author} {\bibfnamefont {H.}~\bibnamefont {Xu}},
  \bibinfo {author} {\bibfnamefont {J.}~\bibnamefont {Berg}},\ and\ \bibinfo
  {author} {\bibfnamefont {E.}~\bibnamefont {Bodenschatz}},\ }\bibfield
  {title} {\bibinfo {title} {The role of pair dispersion in turbulent flow},\
  }\href {https://doi.org/10.1126/science.1121726} {\bibfield  {journal}
  {\bibinfo  {journal} {Science}\ }\textbf {\bibinfo {volume} {311}},\ \bibinfo
  {pages} {835} (\bibinfo {year} {2006})}\BibitemShut {NoStop}%
\bibitem [{\citenamefont {Ouellette}\ \emph {et~al.}(2006)\citenamefont
  {Ouellette}, \citenamefont {Xu}, \citenamefont {Bourgoin},\ and\
  \citenamefont {Bodenschatz}}]{Ouellette2006a}%
  \BibitemOpen
  \bibfield  {author} {\bibinfo {author} {\bibfnamefont {N.~T.}\ \bibnamefont
  {Ouellette}}, \bibinfo {author} {\bibfnamefont {H.}~\bibnamefont {Xu}},
  \bibinfo {author} {\bibfnamefont {M.}~\bibnamefont {Bourgoin}},\ and\
  \bibinfo {author} {\bibfnamefont {E.}~\bibnamefont {Bodenschatz}},\
  }\bibfield  {title} {\bibinfo {title} {An experimental study of turbulent
  relative dispersion models},\ }\href
  {https://doi.org/10.1088/1367-2630/8/6/109} {\bibfield  {journal} {\bibinfo
  {journal} {New J. Phys.}\ }\textbf {\bibinfo {volume} {8}},\ \bibinfo {pages}
  {109} (\bibinfo {year} {2006})}\BibitemShut {NoStop}%
\bibitem [{\citenamefont {Polanco}\ \emph {et~al.}(2018)\citenamefont
  {Polanco}, \citenamefont {Vinkovic}, \citenamefont {Stelzenmuller},
  \citenamefont {Mordant},\ and\ \citenamefont {Bourgoin}}]{Polanco2018}%
  \BibitemOpen
  \bibfield  {author} {\bibinfo {author} {\bibfnamefont {J.~I.}\ \bibnamefont
  {Polanco}}, \bibinfo {author} {\bibfnamefont {I.}~\bibnamefont {Vinkovic}},
  \bibinfo {author} {\bibfnamefont {N.}~\bibnamefont {Stelzenmuller}}, \bibinfo
  {author} {\bibfnamefont {N.}~\bibnamefont {Mordant}},\ and\ \bibinfo {author}
  {\bibfnamefont {M.}~\bibnamefont {Bourgoin}},\ }\bibfield  {title} {\bibinfo
  {title} {Relative dispersion of particle pairs in turbulent channel flow},\
  }\href {https://doi.org/10.1016/j.ijheatfluidflow.2018.04.007} {\bibfield
  {journal} {\bibinfo  {journal} {Int. J. Heat Fluid Flow}\ }\textbf {\bibinfo
  {volume} {71}},\ \bibinfo {pages} {231} (\bibinfo {year} {2018})}\BibitemShut
  {NoStop}%
\bibitem [{\citenamefont {Bitane}\ \emph {et~al.}(2012)\citenamefont {Bitane},
  \citenamefont {Homann},\ and\ \citenamefont {Bec}}]{Bitane2012}%
  \BibitemOpen
  \bibfield  {author} {\bibinfo {author} {\bibfnamefont {R.}~\bibnamefont
  {Bitane}}, \bibinfo {author} {\bibfnamefont {H.}~\bibnamefont {Homann}},\
  and\ \bibinfo {author} {\bibfnamefont {J.}~\bibnamefont {Bec}},\ }\bibfield
  {title} {\bibinfo {title} {Time scales of turbulent relative dispersion},\
  }\href {https://doi.org/10.1103/PhysRevE.86.045302} {\bibfield  {journal}
  {\bibinfo  {journal} {Phys. Rev. E}\ }\textbf {\bibinfo {volume} {86}},\
  \bibinfo {pages} {045302(R)} (\bibinfo {year} {2012})}\BibitemShut {NoStop}%
\bibitem [{\citenamefont {Frisch}(1995)}]{Frisch1995}%
  \BibitemOpen
  \bibfield  {author} {\bibinfo {author} {\bibfnamefont {U.}~\bibnamefont
  {Frisch}},\ }\href {https://doi.org/10.1017/CBO9781139170666} {\emph
  {\bibinfo {title} {Turbulence: {{The Legacy}} of {{A}}.{{N}}.
  {{Kolmogorov}}}}},\ \bibinfo {edition} {1st}\ ed.\ (\bibinfo  {publisher}
  {{Cambridge University Press}},\ \bibinfo {year} {1995})\BibitemShut
  {NoStop}%
\bibitem [{\citenamefont {Falkovich}\ \emph {et~al.}(2001)\citenamefont
  {Falkovich}, \citenamefont {Gaw{\c e}dzki},\ and\ \citenamefont
  {Vergassola}}]{Falkovich2001}%
  \BibitemOpen
  \bibfield  {author} {\bibinfo {author} {\bibfnamefont {G.}~\bibnamefont
  {Falkovich}}, \bibinfo {author} {\bibfnamefont {K.}~\bibnamefont {Gaw{\c
  e}dzki}},\ and\ \bibinfo {author} {\bibfnamefont {M.}~\bibnamefont
  {Vergassola}},\ }\bibfield  {title} {\bibinfo {title} {Particles and fields
  in fluid turbulence},\ }\href@noop {} {\bibfield  {journal} {\bibinfo
  {journal} {Rev. Mod. Phys.}\ }\textbf {\bibinfo {volume} {73}},\ \bibinfo
  {pages} {913} (\bibinfo {year} {2001})}\BibitemShut {NoStop}%
\bibitem [{\citenamefont {Ott}\ and\ \citenamefont {Mann}(2000)}]{Ott2000}%
  \BibitemOpen
  \bibfield  {author} {\bibinfo {author} {\bibfnamefont {S.}~\bibnamefont
  {Ott}}\ and\ \bibinfo {author} {\bibfnamefont {J.}~\bibnamefont {Mann}},\
  }\bibfield  {title} {\bibinfo {title} {An experimental investigation of the
  relative diffusion of particle pairs in three-dimensional turbulent flow},\
  }\href {https://doi.org/10.1017/S0022112000001658} {\bibfield  {journal}
  {\bibinfo  {journal} {J. Fluid Mech.}\ }\textbf {\bibinfo {volume} {422}},\
  \bibinfo {pages} {207} (\bibinfo {year} {2000})}\BibitemShut {NoStop}%
\bibitem [{\citenamefont {Hill}(2006)}]{Hill2006}%
  \BibitemOpen
  \bibfield  {author} {\bibinfo {author} {\bibfnamefont {R.~J.}\ \bibnamefont
  {Hill}},\ }\bibfield  {title} {\bibinfo {title} {Opportunities for use of
  exact statistical equations},\ }\href
  {https://doi.org/10.1080/14685240600595636} {\bibfield  {journal} {\bibinfo
  {journal} {J. Turbul.}\ }\textbf {\bibinfo {volume} {7}},\ \bibinfo {pages}
  {N43} (\bibinfo {year} {2006})}\BibitemShut {NoStop}%
\bibitem [{\citenamefont {Pumir}(1994)}]{Pumir1994}%
  \BibitemOpen
  \bibfield  {author} {\bibinfo {author} {\bibfnamefont {A.}~\bibnamefont
  {Pumir}},\ }\bibfield  {title} {\bibinfo {title} {A numerical study of
  pressure fluctuations in three-dimensional, incompressible, homogeneous,
  isotropic turbulence},\ }\href {https://doi.org/10.1063/1.868213} {\bibfield
  {journal} {\bibinfo  {journal} {Phys. Fluids}\ }\textbf {\bibinfo {volume}
  {6}},\ \bibinfo {pages} {2071} (\bibinfo {year} {1994})}\BibitemShut
  {NoStop}%
\bibitem [{\citenamefont {Naso}\ and\ \citenamefont
  {Godeferd}(2012)}]{Naso2012}%
  \BibitemOpen
  \bibfield  {author} {\bibinfo {author} {\bibfnamefont {A.}~\bibnamefont
  {Naso}}\ and\ \bibinfo {author} {\bibfnamefont {F.~S.}\ \bibnamefont
  {Godeferd}},\ }\bibfield  {title} {\bibinfo {title} {Statistics of the
  perceived velocity gradient tensor in a rotating turbulent flow},\ }\href
  {https://doi.org/10.1088/1367-2630/14/12/125002} {\bibfield  {journal}
  {\bibinfo  {journal} {New J. Phys.}\ }\textbf {\bibinfo {volume} {14}},\
  \bibinfo {pages} {125002} (\bibinfo {year} {2012})}\BibitemShut {NoStop}%
\bibitem [{\citenamefont {Vallefuoco}\ \emph {et~al.}(2018)\citenamefont
  {Vallefuoco}, \citenamefont {Naso},\ and\ \citenamefont
  {Godeferd}}]{Vallefuoco2018}%
  \BibitemOpen
  \bibfield  {author} {\bibinfo {author} {\bibfnamefont {D.}~\bibnamefont
  {Vallefuoco}}, \bibinfo {author} {\bibfnamefont {A.}~\bibnamefont {Naso}},\
  and\ \bibinfo {author} {\bibfnamefont {F.~S.}\ \bibnamefont {Godeferd}},\
  }\bibfield  {title} {\bibinfo {title} {Small-scale anisotropy induced by
  spectral forcing and by rotation in non-helical and helical turbulence},\
  }\href {https://doi.org/10.1080/14685248.2017.1400667} {\bibfield  {journal}
  {\bibinfo  {journal} {J. Turbul.}\ }\textbf {\bibinfo {volume} {19}},\
  \bibinfo {pages} {107} (\bibinfo {year} {2018})}\BibitemShut {NoStop}%
\bibitem [{\citenamefont {Cichowlas}\ \emph {et~al.}(2005)\citenamefont
  {Cichowlas}, \citenamefont {Bona{\"i}ti}, \citenamefont {Debbasch},\ and\
  \citenamefont {Brachet}}]{Cichowlas2005}%
  \BibitemOpen
  \bibfield  {author} {\bibinfo {author} {\bibfnamefont {C.}~\bibnamefont
  {Cichowlas}}, \bibinfo {author} {\bibfnamefont {P.}~\bibnamefont
  {Bona{\"i}ti}}, \bibinfo {author} {\bibfnamefont {F.}~\bibnamefont
  {Debbasch}},\ and\ \bibinfo {author} {\bibfnamefont {M.}~\bibnamefont
  {Brachet}},\ }\bibfield  {title} {\bibinfo {title} {Effective {{Dissipation}}
  and {{Turbulence}} in {{Spectrally Truncated Euler Flows}}},\ }\href
  {https://doi.org/10.1103/PhysRevLett.95.264502} {\bibfield  {journal}
  {\bibinfo  {journal} {Phys. Rev. Lett.}\ }\textbf {\bibinfo {volume} {95}},\
  \bibinfo {pages} {264502} (\bibinfo {year} {2005})}\BibitemShut {NoStop}%
\bibitem [{\citenamefont {Tennekes}\ and\ \citenamefont
  {Lumley}(1972)}]{Tennekes1972}%
  \BibitemOpen
  \bibfield  {author} {\bibinfo {author} {\bibfnamefont {H.}~\bibnamefont
  {Tennekes}}\ and\ \bibinfo {author} {\bibfnamefont {J.~L.}\ \bibnamefont
  {Lumley}},\ }\href@noop {} {\emph {\bibinfo {title} {A {{First Course}} in
  {{Turbulence}}}}}\ (\bibinfo  {publisher} {{MIT Press}},\ \bibinfo {year}
  {1972})\BibitemShut {NoStop}%
\bibitem [{\citenamefont {Baroud}\ \emph {et~al.}(2002)\citenamefont {Baroud},
  \citenamefont {Plapp}, \citenamefont {She},\ and\ \citenamefont
  {Swinney}}]{Baroud2002}%
  \BibitemOpen
  \bibfield  {author} {\bibinfo {author} {\bibfnamefont {C.~N.}\ \bibnamefont
  {Baroud}}, \bibinfo {author} {\bibfnamefont {B.~B.}\ \bibnamefont {Plapp}},
  \bibinfo {author} {\bibfnamefont {Z.-S.}\ \bibnamefont {She}},\ and\ \bibinfo
  {author} {\bibfnamefont {H.~L.}\ \bibnamefont {Swinney}},\ }\bibfield
  {title} {\bibinfo {title} {Anomalous {{Self-Similarity}} in a {{Turbulent
  Rapidly Rotating Fluid}}},\ }\href
  {https://doi.org/10.1103/PhysRevLett.88.114501} {\bibfield  {journal}
  {\bibinfo  {journal} {Phys. Rev. Lett.}\ }\textbf {\bibinfo {volume} {88}},\
  \bibinfo {pages} {114501} (\bibinfo {year} {2002})}\BibitemShut {NoStop}%
\bibitem [{\citenamefont {Mininni}\ and\ \citenamefont
  {Pouquet}(2010)}]{Mininni2010}%
  \BibitemOpen
  \bibfield  {author} {\bibinfo {author} {\bibfnamefont {P.~D.}\ \bibnamefont
  {Mininni}}\ and\ \bibinfo {author} {\bibfnamefont {A.}~\bibnamefont
  {Pouquet}},\ }\bibfield  {title} {\bibinfo {title} {Rotating helical
  turbulence. {{I}}. {{Global}} evolution and spectral behavior},\ }\href
  {https://doi.org/10.1063/1.3358466} {\bibfield  {journal} {\bibinfo
  {journal} {Phys. Fluids}\ }\textbf {\bibinfo {volume} {22}},\ \bibinfo
  {pages} {035105} (\bibinfo {year} {2010})}\BibitemShut {NoStop}%
\bibitem [{\citenamefont {Biferale}\ \emph {et~al.}(2016)\citenamefont
  {Biferale}, \citenamefont {Bonaccorso}, \citenamefont {Mazzitelli},
  \citenamefont {{van Hinsberg}}, \citenamefont {Lanotte}, \citenamefont
  {Musacchio}, \citenamefont {Perlekar},\ and\ \citenamefont
  {Toschi}}]{Biferale2016}%
  \BibitemOpen
  \bibfield  {author} {\bibinfo {author} {\bibfnamefont {L.}~\bibnamefont
  {Biferale}}, \bibinfo {author} {\bibfnamefont {F.}~\bibnamefont
  {Bonaccorso}}, \bibinfo {author} {\bibfnamefont {I.~M.}\ \bibnamefont
  {Mazzitelli}}, \bibinfo {author} {\bibfnamefont {M.~A.~T.}\ \bibnamefont
  {{van Hinsberg}}}, \bibinfo {author} {\bibfnamefont {A.~S.}\ \bibnamefont
  {Lanotte}}, \bibinfo {author} {\bibfnamefont {S.}~\bibnamefont {Musacchio}},
  \bibinfo {author} {\bibfnamefont {P.}~\bibnamefont {Perlekar}},\ and\
  \bibinfo {author} {\bibfnamefont {F.}~\bibnamefont {Toschi}},\ }\bibfield
  {title} {\bibinfo {title} {Coherent {{Structures}} and {{Extreme Events}} in
  {{Rotating Multiphase Turbulent Flows}}},\ }\href
  {https://doi.org/10.1103/PhysRevX.6.041036} {\bibfield  {journal} {\bibinfo
  {journal} {Phys. Rev. X}\ }\textbf {\bibinfo {volume} {6}},\ \bibinfo {pages}
  {041036} (\bibinfo {year} {2016})}\BibitemShut {NoStop}%
\bibitem [{\citenamefont {Zhou}(1995)}]{Zhou1995}%
  \BibitemOpen
  \bibfield  {author} {\bibinfo {author} {\bibfnamefont {Y.}~\bibnamefont
  {Zhou}},\ }\bibfield  {title} {\bibinfo {title} {A phenomenological treatment
  of rotating turbulence},\ }\href {https://doi.org/10.1063/1.868457}
  {\bibfield  {journal} {\bibinfo  {journal} {Phys. Fluids}\ }\textbf {\bibinfo
  {volume} {7}},\ \bibinfo {pages} {2092} (\bibinfo {year} {1995})}\BibitemShut
  {NoStop}%
\bibitem [{\citenamefont {Thiele}\ and\ \citenamefont
  {M{\"u}ller}(2009)}]{Thiele2009}%
  \BibitemOpen
  \bibfield  {author} {\bibinfo {author} {\bibfnamefont {M.}~\bibnamefont
  {Thiele}}\ and\ \bibinfo {author} {\bibfnamefont {W.-C.}\ \bibnamefont
  {M{\"u}ller}},\ }\bibfield  {title} {\bibinfo {title} {Structure and decay of
  rotating homogeneous turbulence},\ }\href
  {https://doi.org/10.1017/S002211200999067X} {\bibfield  {journal} {\bibinfo
  {journal} {J. Fluid Mech.}\ }\textbf {\bibinfo {volume} {637}},\ \bibinfo
  {pages} {425} (\bibinfo {year} {2009})}\BibitemShut {NoStop}%
\bibitem [{\citenamefont {Elsinga}\ \emph {et~al.}(2022)\citenamefont
  {Elsinga}, \citenamefont {Ishihara},\ and\ \citenamefont
  {Hunt}}]{Elsinga2022}%
  \BibitemOpen
  \bibfield  {author} {\bibinfo {author} {\bibfnamefont {G.~E.}\ \bibnamefont
  {Elsinga}}, \bibinfo {author} {\bibfnamefont {T.}~\bibnamefont {Ishihara}},\
  and\ \bibinfo {author} {\bibfnamefont {J.~C.~R.}\ \bibnamefont {Hunt}},\
  }\bibfield  {title} {\bibinfo {title} {Non-local dispersion and the
  reassessment of {{Richardson}}'s $t^3$-scaling law},\ }\href
  {https://doi.org/10.1017/jfm.2021.989} {\bibfield  {journal} {\bibinfo
  {journal} {J. Fluid Mech.}\ }\textbf {\bibinfo {volume} {932}},\ \bibinfo
  {pages} {A17} (\bibinfo {year} {2022})}\BibitemShut {NoStop}%
\bibitem [{\citenamefont {Tan}\ and\ \citenamefont {Ni}(2022)}]{Tan2022}%
  \BibitemOpen
  \bibfield  {author} {\bibinfo {author} {\bibfnamefont {S.}~\bibnamefont
  {Tan}}\ and\ \bibinfo {author} {\bibfnamefont {R.}~\bibnamefont {Ni}},\
  }\bibfield  {title} {\bibinfo {title} {Universality and {{Intermittency}} of
  {{Pair Dispersion}} in {{Turbulence}}},\ }\href
  {https://doi.org/10.1103/PhysRevLett.128.114502} {\bibfield  {journal}
  {\bibinfo  {journal} {Phys. Rev. Lett.}\ }\textbf {\bibinfo {volume} {128}},\
  \bibinfo {pages} {114502} (\bibinfo {year} {2022})}\BibitemShut {NoStop}%
\bibitem [{\citenamefont {Mininni}\ \emph {et~al.}(2012)\citenamefont
  {Mininni}, \citenamefont {Rosenberg},\ and\ \citenamefont
  {Pouquet}}]{Mininni2012}%
  \BibitemOpen
  \bibfield  {author} {\bibinfo {author} {\bibfnamefont {P.~D.}\ \bibnamefont
  {Mininni}}, \bibinfo {author} {\bibfnamefont {D.}~\bibnamefont {Rosenberg}},\
  and\ \bibinfo {author} {\bibfnamefont {A.}~\bibnamefont {Pouquet}},\
  }\bibfield  {title} {\bibinfo {title} {Isotropization at small scales of
  rotating helically driven turbulence},\ }\href
  {https://doi.org/10.1017/jfm.2012.99} {\bibfield  {journal} {\bibinfo
  {journal} {J. Fluid Mech.}\ }\textbf {\bibinfo {volume} {699}},\ \bibinfo
  {pages} {263} (\bibinfo {year} {2012})}\BibitemShut {NoStop}%
\bibitem [{\citenamefont {Delache}\ \emph {et~al.}(2014)\citenamefont
  {Delache}, \citenamefont {Cambon},\ and\ \citenamefont
  {Godeferd}}]{Delache2014}%
  \BibitemOpen
  \bibfield  {author} {\bibinfo {author} {\bibfnamefont {A.}~\bibnamefont
  {Delache}}, \bibinfo {author} {\bibfnamefont {C.}~\bibnamefont {Cambon}},\
  and\ \bibinfo {author} {\bibfnamefont {F.}~\bibnamefont {Godeferd}},\
  }\bibfield  {title} {\bibinfo {title} {Scale by scale anisotropy in freely
  decaying rotating turbulence},\ }\href {https://doi.org/10.1063/1.4864099}
  {\bibfield  {journal} {\bibinfo  {journal} {Phys. Fluids}\ }\textbf {\bibinfo
  {volume} {26}},\ \bibinfo {pages} {025104} (\bibinfo {year}
  {2014})}\BibitemShut {NoStop}%
\bibitem [{\citenamefont {L{\"u}thi}\ \emph {et~al.}(2007)\citenamefont
  {L{\"u}thi}, \citenamefont {Ott}, \citenamefont {Berg},\ and\ \citenamefont
  {Mann}}]{Luthi2007}%
  \BibitemOpen
  \bibfield  {author} {\bibinfo {author} {\bibfnamefont {B.}~\bibnamefont
  {L{\"u}thi}}, \bibinfo {author} {\bibfnamefont {S.}~\bibnamefont {Ott}},
  \bibinfo {author} {\bibfnamefont {J.}~\bibnamefont {Berg}},\ and\ \bibinfo
  {author} {\bibfnamefont {J.}~\bibnamefont {Mann}},\ }\bibfield  {title}
  {\bibinfo {title} {Lagrangian multi-particle statistics},\ }\href
  {https://doi.org/10.1080/14685240701522927} {\bibfield  {journal} {\bibinfo
  {journal} {J. Turbul.}\ }\textbf {\bibinfo {volume} {8}},\ \bibinfo {pages}
  {N45} (\bibinfo {year} {2007})}\BibitemShut {NoStop}%
\bibitem [{\citenamefont {Xu}\ \emph {et~al.}(2008)\citenamefont {Xu},
  \citenamefont {Ouellette},\ and\ \citenamefont {Bodenschatz}}]{Xu2008}%
  \BibitemOpen
  \bibfield  {author} {\bibinfo {author} {\bibfnamefont {H.}~\bibnamefont
  {Xu}}, \bibinfo {author} {\bibfnamefont {N.~T.}\ \bibnamefont {Ouellette}},\
  and\ \bibinfo {author} {\bibfnamefont {E.}~\bibnamefont {Bodenschatz}},\
  }\bibfield  {title} {\bibinfo {title} {Evolution of geometric structures in
  intense turbulence},\ }\href {https://doi.org/10.1088/1367-2630/10/1/013012}
  {\bibfield  {journal} {\bibinfo  {journal} {New J. Phys.}\ }\textbf {\bibinfo
  {volume} {10}},\ \bibinfo {pages} {013012} (\bibinfo {year}
  {2008})}\BibitemShut {NoStop}%
\bibitem [{\citenamefont {Naso}(2019)}]{Naso2019}%
  \BibitemOpen
  \bibfield  {author} {\bibinfo {author} {\bibfnamefont {A.}~\bibnamefont
  {Naso}},\ }\bibfield  {title} {\bibinfo {title} {Multiscale analysis of the
  structure of homogeneous rotating turbulence},\ }\href
  {https://doi.org/10.1103/PhysRevFluids.4.024609} {\bibfield  {journal}
  {\bibinfo  {journal} {Phys. Rev. Fluids}\ }\textbf {\bibinfo {volume} {4}},\
  \bibinfo {pages} {024609} (\bibinfo {year} {2019})}\BibitemShut {NoStop}%
\bibitem [{\citenamefont {Godeferd}\ and\ \citenamefont
  {Lollini}(1999)}]{Godeferd1999}%
  \BibitemOpen
  \bibfield  {author} {\bibinfo {author} {\bibfnamefont {F.~S.}\ \bibnamefont
  {Godeferd}}\ and\ \bibinfo {author} {\bibfnamefont {L.}~\bibnamefont
  {Lollini}},\ }\bibfield  {title} {\bibinfo {title} {Direct numerical
  simulations of turbulence with confinement and rotation},\ }\href
  {https://doi.org/10.1017/S0022112099005637} {\bibfield  {journal} {\bibinfo
  {journal} {J. Fluid Mech.}\ }\textbf {\bibinfo {volume} {393}},\ \bibinfo
  {pages} {257} (\bibinfo {year} {1999})}\BibitemShut {NoStop}%
\end{thebibliography}%

\end{document}